\documentclass[twocolumn, fleqn]{article}
\pdfpageattr{/Group <</S /Transparency /I true /CS /DeviceRGB>>}

\usepackage[svgnames, x11names]{xcolor}
\usepackage[english]{babel}
\usepackage[utf8]{inputenc}

\usepackage{blindtext}
\usepackage{lipsum}
\usepackage{url}
\usepackage{listings}
\usepackage{graphicx}

\usepackage{tikz}
\usetikzlibrary{spy, shapes, backgrounds, fit, angles, quotes, arrows, math, decorations.markings, arrows.meta, decorations.pathmorphing, patterns, fadings, hobby}

\usepackage[europeanresistors, americaninductors, americancurrents, oldvoltagedirection]{circuitikz}

\usepackage{pgfplots}
\pgfdeclarelayer{foreground}
\pgfsetlayers{background,main,foreground}
\DeclareUnicodeCharacter{2212}{−}
\usepgfplotslibrary{groupplots,dateplot}

\usepackage{tcolorbox}
\usepackage{soul}

\usepackage{float}
\usepackage{xcolor}
\usepackage{caption}
\usepackage{subcaption}
\usepackage{dblfloatfix}
\usepackage{xpatch}
\usepackage{appendix}
\usepackage{amsmath,amsfonts,amssymb}
\usepackage{wasysym}
\usepackage{bm}
\usepackage{gensymb}
\usepackage{overpic}

\usepackage{enumitem}

\usepackage{pgfplots}
\pgfplotsset{width=10cm,compat=1.9}
\usepgfplotslibrary{external}

\usepackage{fancyvrb} 

\usepackage{todonotes}
\usepackage{scalerel}
\usepackage{siunitx}
\newcommand{\unit}[1]{\si[sticky-per]{#1}}

\makeatletter
\let\c@author\relax
\makeatother

\usepackage[backend=biber, style=numeric-comp, sorting=none, maxbibnames=10]{biblatex}

\usepackage{csquotes}
\usepackage{pifont}

\usepackage[%
	paperwidth=210mm,
	paperheight=280mm,
	vmargin={19.5mm,18.2mm},
	hmargin={15mm,15mm},
	headsep=12pt,
	footskip=12pt,
	columnsep=18pt
]{geometry}

\usepackage{multirow}

\usepackage{hyperref}

\definecolor{backcolor}{RGB}{240, 240, 240}

\definecolor{S0}{rgb}{1,0.2,0}
\definecolor{S1}{rgb}{1,0.431372549019608,0}
\definecolor{S2}{rgb}{1,0.666666666666667,0}
\definecolor{S3}{rgb}{1,0.901960784313726,0}
\definecolor{S4}{rgb}{0.2,0.8,1}
\definecolor{S6}{rgb}{0.549019607843137,0.450980392156863,1}
\definecolor{S7}{rgb}{0.725490196078431,0.274509803921569,1}
\definecolor{S8}{rgb}{0.901960784313726,0.0980392156862745,1}

\definecolor{CQapp}{RGB}{60, 153, 146}
\definecolor{C_H}{rgb}{0, 1, 0.5}
\definecolor{C_S}{rgb}{0, 1, 0.5}
\definecolor{C_C}{rgb}{0, 1, 0.5}

\makeatletter
\newlength{\fboxrsep}
\newlength{\fboxlsep}
\newlength{\fboxtsep}
\renewcommand\arraystretch{1.5}

\newcommand{\heatCap}{\unit{\joule\per\cubic\metre\kelvin}}

\newcommand{\K}{\unit{\kelvin}}
\newcommand{\mK}{\unit{\milli\kelvin}}

\newcommand{\QGM}{\unit{\watt\tothe{3.4}\per\metre\tothe{5.8}\kelvin}}

\newcommand{\uF}{\unit{\micro\farad}}
\newcommand{\mF}{\unit{\milli\farad}}

\newcommand{\uH}{\unit{\micro\henry}}
\newcommand{\mH}{\unit{\milli\henry}}

\newcommand{\mJ}{\unit{\milli\joule\per\cubic\centi\metre}}
\newcommand{\kW}{\unit{\kilo\watt\per\square\metre}}
\newcommand{\W}{\unit{\watt\per\square\metre}}
\newcommand{\mW}{\unit{\milli\watt}}

\newcommand{\kHz}{\unit{\kilo\hertz}}

\DeclareMathAlphabet\mathbfcal{OMS}{cmsy}{b}{n}

\newcommand{\ecco}{Eccobond\textsuperscript{\textregistered}}
\newcommand{\cx}{Cernox\textsuperscript{\textregistered}}

\newcommand{\Ts}{\ensuremath{T_\text{s}}}
\newcommand{\Tr}{\ensuremath{T_\text{ref}}}

\newcommand{\Tl}{\ensuremath{T_\lambda}}

\newcommand{\QK}{\ensuremath{Q_\text{K}}}
\newcommand{\aK}{\ensuremath{a_\text{K}}}
\newcommand{\nK}{\ensuremath{n_\text{K}}}

\newcommand{\kB}{\ensuremath{k_\text{B}}}

\newcommand{\Qapp}{\ensuremath{Q_\text{app}}}

\newcommand{\aKUnit}{\unit{\watt\per\square\metre\kelvin\tothe{\nK}}}

\newcommand{\etal}{\emph{et al.}}

\newcommand{\sci}[2]{\ensuremath{#1 \cdot 10^{#2}}}

\newcommand{\us}{\unit{\micro\second}}
\newcommand{\ms}{\unit{\milli\second}}

\newcommand{\nm}{\unit{\nano\metre}}
\newcommand{\um}{\unit{\micro\metre}}
\newcommand{\mmm}{\unit{\cubic\metre}}
\newcommand{\mm}{\unit{\square\metre}}

\newcommand{\rhos}{\ensuremath{\rho_\text{s}}}
\newcommand{\rhon}{\ensuremath{\rho_\text{n}}}

\newcommand{\vs}{\ensuremath{\vec{v}_\text{s}}}
\newcommand{\vn}{\ensuremath{\vec{v}_\text{n}}}

\newcommand{\partialf}[2]{\ensuremath{\frac{\partial #1}{\partial #2}}}

\newcommand{\abs}[1]{\ensuremath{\left|#1\right|}}

\newcommand{\sensor}[1]{%
	\mbox{%
		\vphantom{#1}\smash{%
			\tcbox[
			on line,
			boxsep=0pt,
			left=2.5pt, right=2.5pt,
			top=2pt, bottom=2pt,
			colback=LightGray, colframe=White,
			boxrule=0pt,
			arc=1mm, auto outer arc,
			]{\texttt{\textbf{#1}}}
		}%
	}%
	\hspace{-7pt}
}

\makeatletter
\patchcmd{\SOUL@ulunderline}{\dimen@}{\SOUL@dimen}{}{}
\patchcmd{\SOUL@ulunderline}{\dimen@}{\SOUL@dimen}{}{}
\patchcmd{\SOUL@ulunderline}{\dimen@}{\SOUL@dimen}{}{}
\newdimen\SOUL@dimen
\makeatother

\DeclareRobustCommand{\legendEntry}[1]{%
	{\sethlcolor{LightSteelBlue!35!}%
		\hl{\,#1\,}%
	}%
}

\tikzset{
	pointer/.style={
		thick,
		shorten >= 4pt,
		shorten <= 0pt,
		decoration={markings, mark={at position 1 with {\arrow{latex[line width=0.4pt, length=2.5pt,width=4pt]}}}},
		postaction=decorate
	},
	arrow/.style={
		thick,
		shorten >= #1,
		shorten <= 2pt,
		decoration={markings, mark={at position 1 with {\arrow{latex[line width=0.4pt, length=2.5pt,width=4pt]}}}},
		postaction=decorate
	},
	arrow/.default=2pt,
	arrowreversed/.style={
		thick,
		shorten >= #1,
		shorten <= 2pt,
		decoration={markings, mark={at position 0 with {\arrow{latex[line width=0.4pt, length=2.5pt,width=4pt]}}}},
		postaction=decorate
	},
	arrowreversed/.default=2pt,
	dot/.style={
		thick,
		shorten >= 2pt,
		shorten <= 2pt,
		decoration={
			markings, mark={at position 1 with {\draw circle [radius=2pt];}}
		},
		postaction=decorate
	},
	dimen/.style={
		thick,
		shorten >= 4pt,
		shorten <= 4pt,
		decoration={
			markings, 
			mark = {
				at position 0 with {
					\arrowreversed{latex[line width=0.4pt, length=2.5pt,width=4pt]|[width=5mm]}
				}
			},
			mark = {
				at position 1 with {
					\arrow{latex[line width=0.4pt, length=2.5pt,width=4pt]|[width=5mm]}
				}
			}
		},
		postaction=decorate
	},
	dimenShort/.style={
		thick,
		shorten >= 4pt,
		shorten <= 4pt,
		decoration={
			markings, 
			mark = {
				at position 0 with {
					\arrowreversed{latex[line width=0.4pt, length=2.5pt,width=4pt]|[width=5mm]}
				}
			},
		},
		postaction=decorate
	},
	declare function={
		atan3(\a,\b)=ifthenelse(atan2(0,1)==90, atan2(\a,\b), atan2(\b,\a));
	},
	kinky cross radius/.initial=+.125cm,
	@kinky cross/.initial=+,
	kinky crosses/.is choice,
	kinky crosses/left/.style={@kinky cross=-},
	kinky crosses/right/.style={@kinky cross=+},
	kinky cross/.style args={(#1)--(#2)}{
		to path={
			let \p{@kc@}=($(\tikztotarget)-(\tikztostart)$),
			\n{@kc@}={atan3(\p{@kc@})+180} in
			-- ($(intersection of \tikztostart--{\tikztotarget} and #1--#2)!%
			\pgfkeysvalueof{/tikz/kinky cross radius}!(\tikztostart)$)
			arc [ radius     =\pgfkeysvalueof{/tikz/kinky cross radius},
			start angle=\n{@kc@},
			delta angle=\pgfkeysvalueof{/tikz/@kinky cross}180 ]
			-- (\tikztotarget)
		}
	}
}

\pgfplotsset{%
	grid style={
		line width=0.1pt, gray!50!
	},
	every axis/.append style={
		label style={font=\small},
		tick label style={font=\small}  
	}
}

\makeatletter
\renewcommand{\paragraph}{%
	\@startsection{paragraph}{4}%
	{\z@}{1mm \@plus 1ex \@minus .2ex}{-1em}%
	{\normalfont\normalsize\bfseries}%
}
\makeatother

\begin{document}	
	\title{Millisecond Time--Scale Measurements of\\Heat Transfer to Confined He~II}
	
	\author{
		Jonas Blomberg Ghini$^{1,2,\ast}$ \and Bernhard Auchmann$^{2}$ \and Bertrand Baudouy$^{3}$
	}
	\date{%
		$^{1}$Department of Physics, Norwegian University of Science and Technology, NTNU, Norway\\%
		$^{2}$European Organization for Nuclear Research, CERN, Switzerland\\%
		$^{3}$Irfu, CEA, Université Paris--Saclay, F-91191 Gif--sur--Yvette, France\\%
		$^{\ast}$Corresponding author: \emph{jonas.blomberg.ghini@ntnu.no}%
	}
	\maketitle
	
	\begin{abstract}
%

We explore transient heat transfer, on the millisecond time--scale, from a narrow, rectangular stainless steel heater cooled from one side by He~II confined to a channel of 120~\um\ depth. The helium is isolated from the external bath with the exception of two pin--holes of cross section about 10\%\ that of the channel.

We measure the temperatures of both the heater strip and the channel helium during slow--pulse heating that reaches peak power after 9~\ms, fast--pulse heating that reaches peak power after 100~\us, and step heating that reaches steady power after 100~\us.

Using the steady state Kapitza heat transfer expression at the interface between heater and helium, and the Gorter--Mellink heat transfer regime in the helium channel, we obtain excellent agreement between simulation and measurement during the first 5~\ms\ of slow--pulse tests. Using instead the measured helium temperature in the Kapitza expression, we obtain excellent agreement between the simulated and measured heater response during the first 150~\ms\ of slow--pulse tests.

The same model fails to explain the fast--pulse transient response of the heater and helium, while it can explain the helium response to a step in applied power but not the heater response. The steady state Kapitza expression may therefore not be applicable to heating events that are over within a single millisecond.
	\end{abstract}
	
	\section{Introduction}

The Large Hadron Collider (LHC) is subject to a recurrent beam loss event called an Unidentified Falling Object (UFO)~\cite{UFOsInTheLHC_BaerEtal_2012}, happening on the order of 10 to 30 times per hour of LHC operation~\cite[Fig. 2]{ufoIS_and_hoursAfterQuench}. A UFO event is an interaction between the particle beam and dust or debris that is assumed to fall from the top of the LHC beam pipe into the path of the beam, causing inelastic collisions between the beam particles and the dust~\cite[Sec. 3.3.1]{scott_thesis}. The interaction leads to energy deposition in the superconducting magnet that surrounds the beam pipe. This transient energy deposition has an asymmetric Gaussian shape, and the entire UFO event is typically over after just 1~\ms~\cite[Fig. 3.7]{scott_thesis}. If the energy deposition is sufficiently large, the superconducting magnet will quench, meaning it loses superconductivity (locally at first, and then this initial normal--conducting zone propagates to the rest of the magnet). In the case of a magnet quench it will take about 12 hours before normal LHC operation is restored~\cite{TimeLostForBeamDump}. If a that would cause a quench UFO is detected early enough, triggering a beam dump, it still takes about 3 hours to resume normal operation. 

During LHC operation successful mitigation strategies are currently employed such that quenches are avoided, and such that unnecessary beam dumps are kept to a minimum~\cite{HowToSurviveUFO}. In the future, however, the LHC is scheduled for an upgrade to increase the beam energy~\cite{HE_LHC}. For higher beam energy, each UFO event will cause larger losses, because the individual beam--particles impact the UFO with more energy~\cite[Fig. 5.11]{Baer_thesis}. This leads to the need for better understanding of the millisecond time--scale losses UFOs represent.

After analysis of a dedicated beam--induced quench test performed in 2011~\cite{orbitBump_2011}, Bernhard \etal\ found that for transient losses on the millisecond time--scale, there is as much as a factor 4 difference between how much energy is actually necessary to quench a magnet and how much energy an electro--thermal model of the magnet predicts should be sufficient~\cite[Tab. V]{bernhardsBeamInducedPaper_withFactor4}; the magnets are more resilient against quenches than assumed. The model (see Ref.~\cite{QP3_manual}) used to predict the energy necessary to quench simulates a single strand of the magnet cable cooled by He~II in contact with the strand surface, and subjects the strand to the losses measured during the quench test. Helium cooling is accounted for by using a steady state Kapitza heat transfer relation at the interface between the strand and surrounding helium. This approach is typical of magnet stability modelling~\cite{Arjan_CUDI, BreschiGoodCableModellingResults}. 

Finding this discrepancy between simulated and observed behaviour lead to the interest in measuring the temperature response of a heater cooled by confined helium during millisecond time--scale losses. Section \ref{sec:theory} expands on both theory and background from previous experimental work. An important shortcoming of current understanding is that the standard Kapitza surface heat transfer model was developed for steady state heating into large volumes of helium and we need to ensure its validity for use during transient heating into a confined volume of helium. There is also very little published data available from millisecond time--scale measurements of heat transfer to confined volumes, and our work will help fill this knowledge gap.

The measurement campaign presented in this paper revolves around two main time--profiles for applied heating power; 1) we start with a slow pulse that reaches peak power after about 9~\ms, with a long tail, lasting a total of 400 to 500~\ms. This helps assess the validity of using the Kapitza surface heat transfer expression in the confined channel helium volume. 2) to approximate UFO events we use fast pulses that deliver peak power after around 100~\us, that last a total of 800 to 1000~\us. We also apply steps to the heater, which reach their steady state power within about 100~\us\ as a means to compare closed channel and open bath test results.

The open bath test results are discussed in--depth in a separate publication (Ref.~\cite{OpenBathPaper}). For convenience, when we refer to the open bath paper, which shares many similarities in the setup and measurement procedure, we will refer to it as the Open Bath Paper.
		
	\section{Theory and Background} \label{sec:theory}
		In this paper we we consider heat transfer from a narrow rectangular heater strip exposed on one side to helium along its entire length. The helium is confined to a shallow channel. Heat transfer across the heater--helium interface is governed by the Kapitza heat transfer expression in Equation~\eqref{eq:kapitza}, which was first proposed by Claudet and Seyfert~\cite[Eq. 1]{claudet_seyfert_standardKapitzaExpression};
\begin{equation} \label{eq:kapitza}
	Q_\text{K} = \aK\ \left( \Ts^{\nK} - \Tr^{\nK} \right),
\end{equation}
where \QK\ is the Kapitza heat flux, \Ts\ is the temperature of the heater at the heater--helium interface, \Tr\ is a reference temperature equal to that of the helium in the channel, and \aK\ and \nK\ are two fit parameters. In our Open Bath Paper we determine the Kapitza parameters for our stainless steel heater strip to be \aK\ = 1316.8~\aKUnit\ and \nK\ = 2.528. For convenience, when we refer to the surface temperature of the heater, we always mean the temperature at the interface between heater and helium, \Ts.

Heat transfer to He~II in the Kapitza regime involves the transmission of thermal phonons across the heater--helium interface~\cite[Sec. 7.5]{VanSciver}\cite[Chap. 23]{Khalatnikov_book}. The dominant phonon wavelength excited in the helium are approximately $\lambda = h v_\text{He} / 3.8 \kB\ T$, where $h$ is the Planck constant, $v_\text{He}$ the speed of sound in helium, \kB\ is the Boltzmann constant, and $T$ is the helium temperature~\cite[p. 168]{ThermalPhononWavelength}. For temperatures between 1.9~\K\ and \Tl\ (= 2.165~\K), taking the speed of sound to be around 200~\unit{\metre\per\second}, we get a phonon wavelength on the order of 1~\nm. Compared with the channel we use, of 120~\um\ depth, a phonon transmitted across the heater--helium interface will not distinguish the channel from an open bath. As such, we expect the Kapitza expression to hold, at least for steady state, also in the confined geometry.

Katerberg and Anderson verified that the measured Kapitza resistance between He~II and a copper heater remained the same when measured during steady applied heat or during oscillating heating power~\cite{KaterbergAnderson_DC_AC_kapitzaResistance}. Their fastest heating was generated by a sinusoidal 600~\unit{\hertz} current applied to the heater, leading to a peak in applied heating power every half--cycle, or 830~\us. They did not find conclusive evidence of a frequency dependence of the Kapitza resistance, though, as they point out, above about 1.6~\K\ their data is not reproducible, and there seems to be a weak tendency for the Kapitza resistance to be lower for higher frequency. From this, we expect the Kapitza expression to certainly be valid for the applied heating that peaks after 9~\ms, but it is not clear if it remains valid down to heating that peaks after just 100~\us.

\subsection{Millisecond Time--Scale Data}
	Transient heat transfer experiments in He~II tend to fall into three main categories;
	\begin{enumerate}
		\item
			Tests designed to assess the time it takes before onset of film boiling after subjecting a heater submerged in an open bath to a very strong heat flux, usually in excess of 100~\kW, depending on the bath temperature (see the Open Bath Paper for detailed discussion~\cite{OpenBathPaper}). The time--scale of these measurements are often on the order of 100~\us\ or less, but they do not consider the time during which the heater warms up, and how the Kapitza expression may describe this time window.
		\item 
			Tests designed to assess propagation of second sound waves in He~II (Ref.~\cite{NemirovskiiAndTsoi_transientRegimeClassification} discusses several such tests). The applied heating power densities of such tests well in excess of 100~\kW\ and at time--scales on the order of 10 to 100~\us. The behaviour of the heater exciting these pulses is generally not considered.
		\item 
			Tests designed to assess lower heat fluxes propagating along a channel of He~II~\cite{GorterAndMellink_firstMutualFriction, Vinen_series_II, Chase_geometry, VanSciver_transientInLongCoiledPipe, Vitrano_2020}. While the applied heating power densities of such tests are often in the range of a few tens of \kW, the transient data, if at all available, is usually on the 0.1~s time--scale or slower, because heat propagation along even relatively short channels of He~II is quite slow.
	\end{enumerate}

	So, there is a gap in available data; for applied heating power density on the order of 1 to 100~\kW, there are no published results nn the millisecond time--scale that considers both helium and heater behaviour.

\subsection{Heat Transfer in Confined Helium Geometries} \label{sec:steadyStateInChannel}
	Several experiments have been done in the past mapping out the steady state heat transfer characteristics of helium in varying sorts of confined geometries. Warren and Caspi measured the heater surface temperature of a cylindrical heater exposed on one side to a He~II filled gap of varying thickness\cite{WarrenAndCaspi_narrowGap}; Chen and Van~Sciver built a rectangular channel 127~mm long and 12.7~mm wide, measuring the heater surface and channel helium temperatures for varying channel depths and angles of inclination\cite{ChenAndVanSciver_orientationDependence_ChannelLikeMine}; Kobayashi \etal\ built a short channel, open on either end to a large bath of He~II, where they measured the heater surface temperature, the channel helium temperature immediately adjacent to the heater, and the temperature of the helium adjacent to the wall opposite the heater\cite{Kobayashi_subcooledHeILayer}; Granieri conducted a large study of steady state heat transfer in Rutherford cables of the type used in the LHC main dipoles\cite[Chap. 3]{Granieri_thesis}.
	
	They all find steady state results roughly characterised by four regimes;
	\begin{enumerate}
		\item
			Below a small critical heat flux $Q_\lambda$, for which the helium temperature remains below \Tl, no thermal gradients are observed within the confined volume. Heat is applied to the confined helium according to the Kapitza regime, and it is transferred out of the helium by the Gorter--Mellink regime (see Section \ref{sec:heatTransferInHelium});
		\item
			Between $Q_\lambda$ and an intermediate heat flux $Q_\text{NucBoil}$, relatively poor heat transfer takes place between the heater and the helium, and most of the applied heat is transferred by conduction through the solids in the setup, rather than to the helium itself. This is recognised as a natural convection--like regime;
		\item 
			Between $Q_\text{NucBoil}$ and a higher $Q_\text{FilmBoil}$, strong heat transfer from heater to helium takes place, identified as the nucleate boiling regime. The presence of this regime appears to depend not only on the channel orientation, but also the helium bath temperature. Near \Tl, the regime is seen regardless of channel orientation, while at 1.9~\K\ it is hardly noticeable in some horizontal channels;
		\item 
			Above $Q_\text{FilmBoil}$, film boiling develops fully, and, like for the natural convection regime, only poor heater to helium heat transfer takes place.
	\end{enumerate}

	While these critical heat fluxes are defined for steady state heating, which we are not investigating, we expect to see evidence of the transition into natural convection for tests where we heat the channel helium up to \Tl.

	\subsubsection{Transient Heat Transfer to a Channel}
		Okamura \etal\ measured transient temperatures in a 170~mm long, 7~mm wide channel of varying depth, open on either end, subject to steps in applied heating power density\cite{Okamura_channelExperiment_verySimilarToMyStuff}. Their steady state results are not entirely in line with the expected four regimes; in particular there is no clear natural convection regime separating a low--heat flux Kapitza regime from a high--heat flux nucleate boiling regime. Furthermore, when their channel is horizontal, with the heater facing upwards, there is no nucleate boiling regime clearly present. However, they show transient measurements on the 0.1~second time scale. The key result is that the final steady state for $T_\text{channel} < \Tl$ appears to need on the order of 3 to 5 seconds to be established.
		
		Their results also show behaviour analogous to that seen in open baths or long channels heated from one end; for an applied heating power density larger than $Q_\lambda$, there is a finite life--time during which strong Kapitza cooling persists before the confined helium volume reaches \Tl, and there is a heat transfer regime change.

\subsection{Heat Transfer in Helium} \label{sec:heatTransferInHelium}
	Above \Tl, in a horizontal channel filled with He~I, where we neglect convective effects, the thermal conductivity of the helium is rather low, and comparable to that of an insulating material like \ecco, for instance. Below \Tl, however, the effective thermal conductivity of the same channel, now filled with He~II, becomes more akin to, or even higher than, that of high--purity copper. The high effective thermal conductivity of He~II is due to the Gorter--Mellink mutual friction regime. There exists also a laminar regime in He~II. The Gorter--Mellink regime is dominant under two main conditions; 1) turbulence must be fully developed; and 2) there must be zero net mass flow of He~II. Vinen made the early key contributions to understanding of turbulence in He~II, finding that for an applied heating power density \Qapp, it takes,
	\begin{equation} \label{eq:turbulenceOnsetTime}
		\tau = a {\Qapp}^{-3/2},
	\end{equation}
	seconds to develop turbulence, where $a$ is a geometry--dependent fit parameter~\cite{Vinen_series_II}. Chase found that in channels with obstructing orifices the time to develop turbulence becomes essentially zero, giving $a$ = 0, while $a$ is otherwise on the order of 100~\unit{\kilo\second\watt\tothe{3/2}\per\cubic\metre} around 1.9~\K, falling towards zero as the temperature approaches \Tl\cite{Chase_geometry}. 
	
	There isn't a neat criterion to assess whether or not the zero--net--mass flow condition is met. Heat transfer in He~II is associated with fluid flow, and is described by Landau's two--fluid model\cite{Landau_excitationSpectrum}. Zero net mass flow means, in the two--fluid picture, that the normal--fluid component momentum flux $\rhon\vn$ exactly balances the superfluid component momentum flux $\rhos\vs$. In our work, we assume the zero--net--mass flow (counter--flow) condition to be met at all times, meaning we assume the Gorter--Mellink regime is always dominant.
	
	In the Gorter--Mellink regime, the local helium heat flux $Q_\text{GM}$ is given as;
	\begin{equation} \label{eq:gorterMellinkHeatFlux_withVariableExponent}
		Q_\text{GM} = \left[-f^{-1}(T) \nabla T\right]^{1/m},
	\end{equation}
	where $f^{-1}$ is the thermal conductivity function of He~II (where we use that proposed by Sato \etal\ \cite{Sato_fit}), and $m$ is an exponent originally found by Gorter and Mellink to be 3\cite{GorterAndMellink_firstMutualFriction}, while Sato \etal\ later propose $m = 3.4$\cite{Sato_exponent} (which we use herein).
		
	Using this, a modified heat equation is obtained from Fourier's law of cooling to express an effective thermal conductivity from Equation \ref{eq:gorterMellinkHeatFlux_withVariableExponent};
	\begin{equation} \label{eq:transientHeII}
		\rho(T)C_\text{p}(T) \partialf{T}{t} = \nabla \left\{ \left[-f^{-1}(T) \frac{1}{\abs{\nabla T}^{n-1}}\right]^{1/n} \nabla T \right\},
	\end{equation}
	where $C_\text{p}$ is the specific heat capacity. This assumes that Equation \ref{eq:gorterMellinkHeatFlux_withVariableExponent}, which, strictly speaking, is a steady state expression, also applies under transient conditions.
	
	Seyfert \etal\ used this kind of approach to model heat transfer in an annular channel\cite{SeyfertEtal_concentricCylindersTimeToBoilingAndTimeDependentSimulation}. Okamura \etal\ did the same for He~II in a pipe subject to slowly varying applied heating power\cite{Okamura_sinusoidHeatTransferAtMouthOfPipe}. Fuzier and Van~Sciver modified Equation \eqref{eq:transientHeII} to account for forced flow of He~II in a pipe heated at one end, and found good agreement in the limiting cases of; 1) large mass flow, where thermal convection dominated heat transfer; and 2) low mass flow, where Gorter--Mellink counterflow dominates heat transfer\cite{FuzierAndVanSciver_forcedFlowExperiment}\cite[Sec. 5]{fuzierThesis}.
		
%
		
	\paragraph{Helium material properties}
		We take helium material properties from Arp~\etal~\cite{HeliumProperties_ArpMcCartyFriend}. We use their data to build cubic spline interpolation functions between 1.7 and 100~\K.
		
	\section{Setup}
\begin{figure}[ht]
	\centering
	\includegraphics{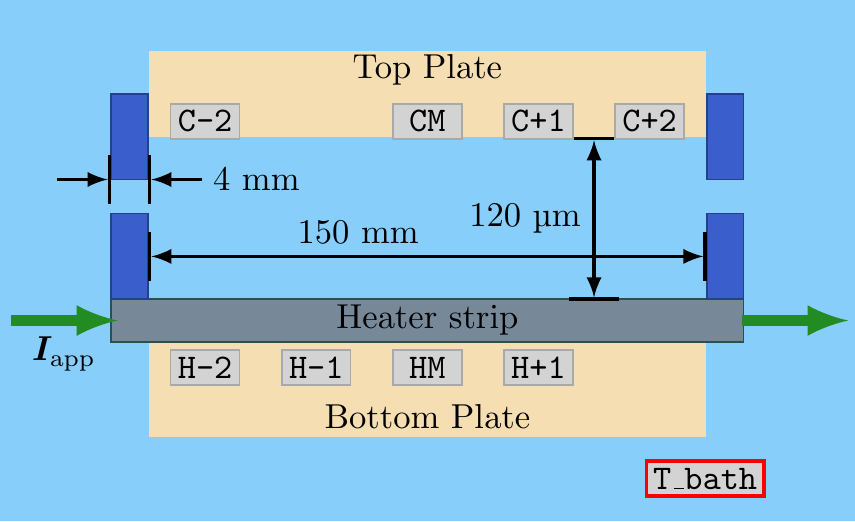}
	\caption{Diagram of the closed channel setup used for experiments. \sensor{H} sensors measure the heater temperature, and \sensor{C} sensors measure the channel helium temperature. \sensor{T\_bath} is the reference probe. The channel is 150~mm long and 120~\um\ deep. It is 3.1~mm wide, while the heater is 3~mm wide. The pin--holes at either end are 4~mm long.}
	\label{fig:DiagramOfSample}
\end{figure}

		Most of the experimental setup was described in the Open Bath Paper, and here we will only point out the key differences. Figure \ref{fig:DiagramOfSample} shows a diagram of the experimental sample placed in the cryostat. The \legendEntry{Bottom Plate} is the same plate as the downwards facing heater in the Open Bath Paper, corresponding to the \sensor{D}--sensors in the paper. The \legendEntry{Top Plate} is used to confine the helium to a 120~\um\ deep channel along the heater strip. The two PEEK plates are kept apart by Kapton tape laid along either long--side of the channel, and 16 aluminium bolts (eight per side) clamp the two plates together. Sensors measuring the heater temperature are labelled \sensor{H}, while those measuring the channel helium temperature are labelled \sensor{C}.

The dark blue rectangles at either end of the channel in the figure represent the \ecco\ through which pin--holes are made. The pin--holes have, roughly, a half--circular cross section, with their origins at the middle of the 3.1~mm wide channel groove in the \legendEntry{Top Plate}. Their cross section is \sci{3.405}{-8}~\mm, or 9.15\%\ of the channel cross section. They are 4~mm long, and along their bottoms they are exposed to the bare heater strip. The pin--holes limit the channel length to 150~mm.

\begin{figure}[ht]
	\centering
	\includegraphics{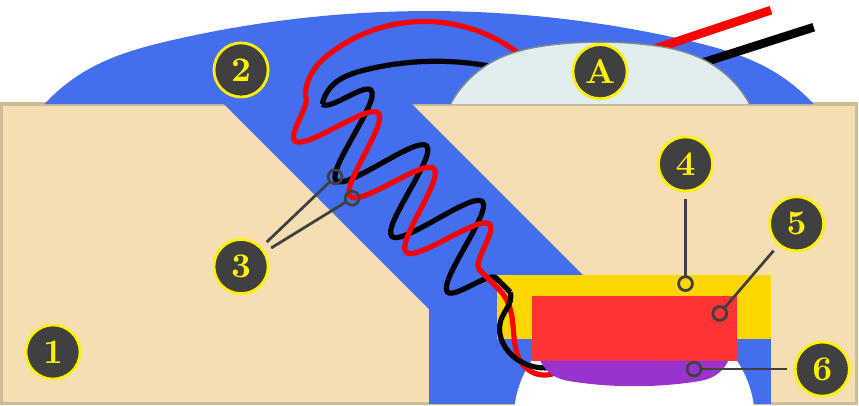}
	\caption{Schematic representation of the region around the \cx\ temperature sensors of the Top Plate, sensing the channel helium temperature. Label (1): Glass--fibre filled PEEK. (2) \ecco. (3): copper sensor lead wires. (4): GE 7031 varnish. (5): sapphire sensor substrate. (6): EPO--TEK H20E epoxy. (A): Soldering point where thin sensor leads join larger sensor lead attachments.}
	\label{fig:SchematicAroundTopPlateSensors}
\end{figure}

Figure \ref{fig:SchematicAroundTopPlateSensors} represents the region around the helium channel sensors. The assembly is essentially the same as for the heater sensors, described in the Open Bath Paper, but there is no Kapton or steel in the Top Plate, and the sensors have been turned so that the sensitive zirconium oxynitride film faces the helium to ensure the fastest possible thermal response time. The labels in the figure refer to the following;
\begin{itemize}
	\itemsep-1pt
	\item[\textbf{(1)}]
		Glass--fibre filled PEEK;
	\item[\textbf{(2)}]
		\ecco\ epoxy used to fill in holes, as well as seal the sample to prevent helium to escape through the crack where the two PEEK plates mate;
	\item[\textbf{(3)}]
		Copper sensor leads, attached by manufacturer;
	\item[\textbf{(4)}]
		GE 7031 varnish used to attach \cx\ sensors in recesses in the Top Plate;
	\item[\textbf{(5)}]
		\cx\ sapphire sensor substrate. Note that the sensors are not entirely flush with the PEEK surface in order to make sure the EPO--TEK beads could not touch the heater strip during testing of the assembly and final mounting;
	\item[\textbf{(6)}]
		EPO--TEK H20E silver filled epoxy used by Lake Shore Cryotronics to attach leads to sensor body;
	\item[\textbf{(A)}]
		Soldering point joining the sensor leads to larger external lead attachments, acting as thermal anchoring for the \cx\ sensors.
\end{itemize}
Note that unlike for the Bottom Plate, where the sensor is embedded within a stack of materials under the heater strip, the sensors on the channel side are mounted so their sensitive part is in direct contact with the helium. This means that the helium sensor temperature measurements do not need a post--processing step to account for a material stack like we do for the heater sensors. Analysis done in this paper uses the same modelling approach for the heater sensors as that used when discussing open bath measurements in our Open Bath Paper. The dimensions we use as a reference for the material stack under the heater strip are the same as well; 50~\um\ of stainless steel for the heater strip, 35~\um\ of varnish used to attach the heater sensor to the heater strip, 200~\um\ of sapphire representing the bulk of the \cx\ sensor, 20~\um\ of EPO--TEK used to attach the copper leads to the sensors, and 20~mm of copper for the leads that run to the back of the sample. Material parameters are discussed in Appendix A in our Open Bath Paper.

\subsection{Calibration}
	The \cx\ sensors are calibrated in--situ against the reference probe in the same way as was done in the Open Bath Paper. The total estimated measurement uncertainty for the closed channel tests presented here is given in Table \ref{tab:uncertainty}.
	
	
	\begin{table}[ht]
		\vspace{-5pt}
		\renewcommand{\arraystretch}{1}
		\caption{Estimated measurement uncertainty $\Delta T$.}
		\label{tab:uncertainty}
		\vspace{-6pt}
		\centering
		\begin{tabular}{r c l r}
			\multicolumn{3}{c}{Range, [\K]} & $\pm\Delta T$, [\mK]	\\ \hline
			1.8		&	---		& \Tl		& 5 \\
			\Tl		&	---		& 2.3		& 19 \\
			2.3		&	---		& 3			& 7 \\
			3		&	---		& 4			& 13 \\
			4		&	---		& 6			& 8 \\
			6		&	---		& 20		& 15
		\end{tabular}
		\vspace{-10pt}
	\end{table}

	\paragraph{Instrumentation}
		There are no differences in instrumentation from that used in the Open Bath Paper, save for two added measurement channels. One is for the extra temperature sensor, as we here have eight in total, rather than seven, and the other added channel is for a synchronisation measurement where the temperature data acquisition system senses the triggering signal to the powering circuit that delivers current to the heater strip. The temperature sensor data acquisition frequency is 20~\kHz\ per sensor/channel.
		
	\paragraph{Post--processing of transient measurements}
		In order to obtain the transient temperature data from our measurements, we compensate for the presence of a filtering capacitor on the \cx\ sensor current excitation sources. This method is described in our Open Bath Paper.

\subsection{Sealing the Helium Channel}
\begin{figure}[ht]
		\centering
		\includegraphics[width=\columnwidth, clip, trim=140 230 220 300]{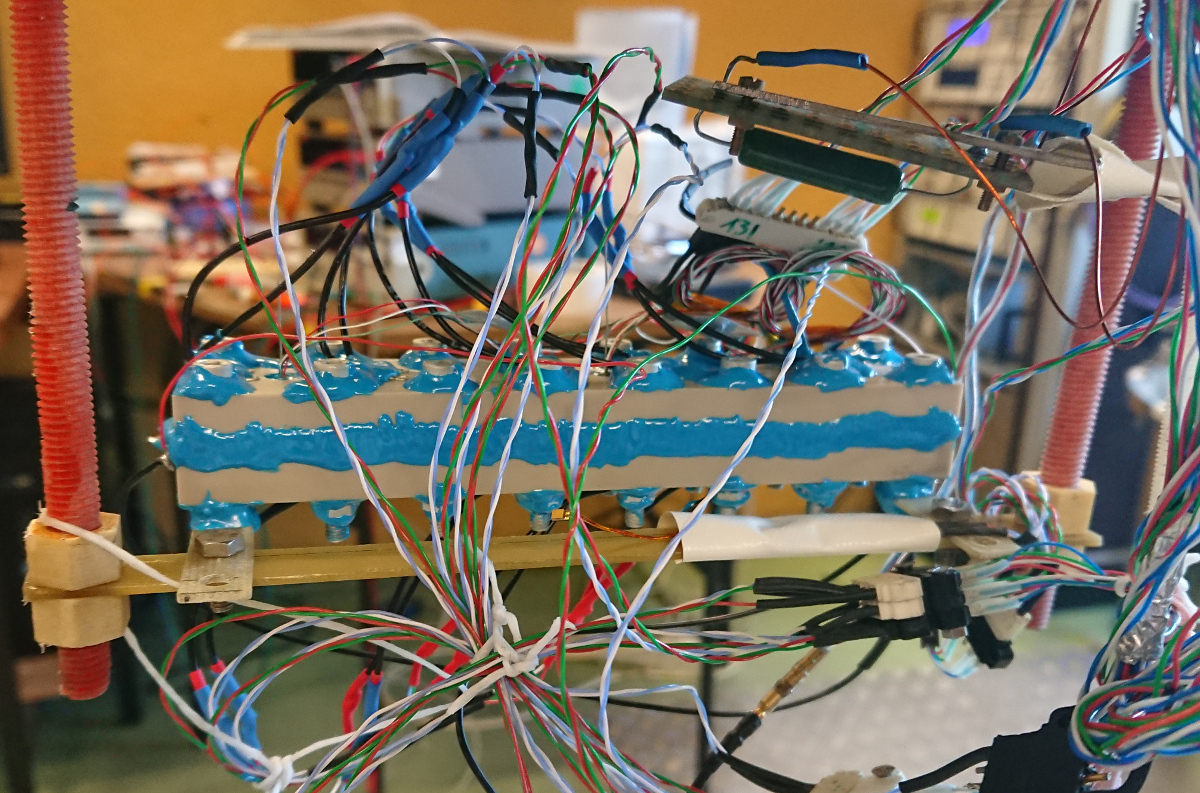}
		\caption{Photograph of setup after sealing gaps with \ecco\ to ensure no continuous helium channels exist between the channel and the outside bath, aside from the pin--holes.}
		\label{fig:Cryostat_Insert_ClosedChannel}
	\end{figure}

	After bolting the two PEEK plates together to confine the helium near the heater strip to the desired channel dimensions, we sealed all remaining gaps, except for the pin--holes, with \ecco, in order to ensure the helium was only in direct thermal contact with the bath through the two pin--holes. Figure \ref{fig:Cryostat_Insert_ClosedChannel} shows a photograph of the setup after \ecco\ sealing.  
	
	Between the two PEEK plates we placed a strip of Kapton tape on either side of the channel. This serves two purposes; 1) the tapes act as a spacer to give the desired channel depth of 120~\um, and 2) the Kapton tapes cover the 16 bolt holes, and the aluminium bolts pass through the tape. So, after tightening the bolts, the layer of tape helps provide a seal to ensure the volumes between the bolt--hole walls and the bolts themselves are completely isolated from the helium in the channel.

\subsection{Measurement Procedure}	
	For our heating tests we use three different types of applied heating power densities;
	\begin{itemize}
		\item
			Steps that reach flat top power after about 100~\us. Applied heating power density reaches 970~\W.
		\item
			Slow RLC--like pulses that reach peak power after around 9~ms, lasting a total of 400 to 500~ms. Peak applied heating power density reaches 4300~\W.
		\item 
			Fast RLC--like pulses that reach peak power after around 100~\us, lasting a total of 800 to 1000~\us. Peak applied heating power density reaches 130~\kW. 
	\end{itemize}

\begin{figure}[ht]
		\centering
		\includegraphics{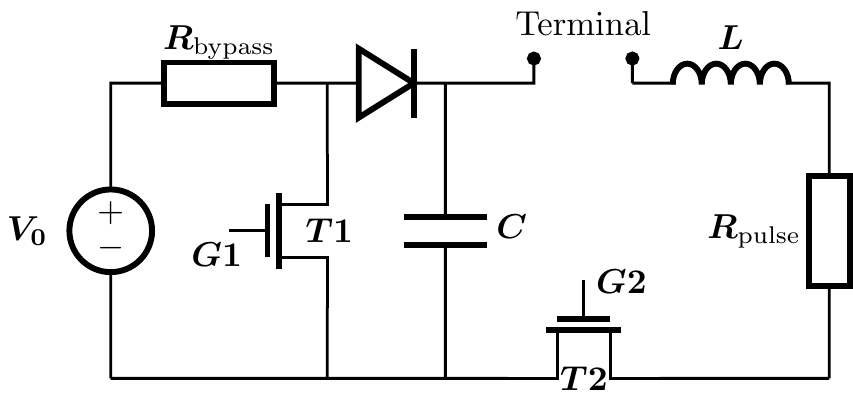}
		\vspace{-10pt}
		\caption{Circuit used to generate the slow and fast RLC--like heating pulses. \legendEntry{$R_\text{pulse}$}, \legendEntry{$L$}, and \legendEntry{$C$} are used to shape the pulse. The transistor \legendEntry{T2} releases energy into the heater strip upon switching \legendEntry{G2}.}
		\label{fig:RLCCircuit_tikz}
	\end{figure}

	To generate the RLC--like pulses we use the circuit shown in Figure \ref{fig:RLCCircuit_tikz}. The circuit components \legendEntry{$R_\text{pulse}$}, \legendEntry{$L$}, and \legendEntry{$C$} are used to shape the pulse. For slow pulses we use $C$ = 40~\mF, $L$ = 8.5~\mH, and $R_\text{pulse}$ = 3.8~\unit{\ohm}. For fast pulses we use $C$ = 160~\uF, $L$ = 93~\uH, and $R_\text{pulse}$ = 1.3~\unit{\ohm}. We connect the heater strip power leads to the \legendEntry{Terminal}, and use the transistor \legendEntry{$T2$} to release the energy stored in the capacitor by switching the gate signal \legendEntry{$G2$}. This switching is what triggers the heater strip data acquisitions system to log data. 
	
	$R_\text{bypass}$, $T2$, and the diode lets us keep the power supply on between tests and avoid any backwards energy flow from the capacitor during the pulse discharge.
	
	The shape of the fast pulses is chosen to be similar to the transient UFO--losses seen in the LHC, while the slow pulses represent a transient that helps determine the region of validity of our modelling efforts.
	
	The heater strip, which is the same as the downwards facing heater from the Open Bath Paper, has a resistance of 0.458~\unit{\ohm}, and to represent the heating power developed in the strip, we use the applied heating power density calculated from the measured heater strip voltage;
	\begin{equation} \label{eq:Qapp}
		\Qapp = \frac{{V_\text{meas}}^2}{R_\text{strip}} \frac{d_\text{strip}}{v_\text{strip}},
	\end{equation}
	with $V_\text{meas}$ the measured voltage, $R_\text{strip}$ = 0.458~\unit{\ohm}, $d_\text{strip}$ = 50~\um, and $v_\text{strip}$ = 150~mm $\times$ 3~mm $\times$ 50~\um, the total volume of the heater strip.
	
	The helium--wetted area of the heater strip is 150~mm $\times$ 3~mm. The helium in the 4~mm long pin--holes also contact the heater.

	\section{Slow Pulses in Applied Heating Power Density} \label{sec:slow}
\begin{figure}[ht]
	\centering
	\includegraphics{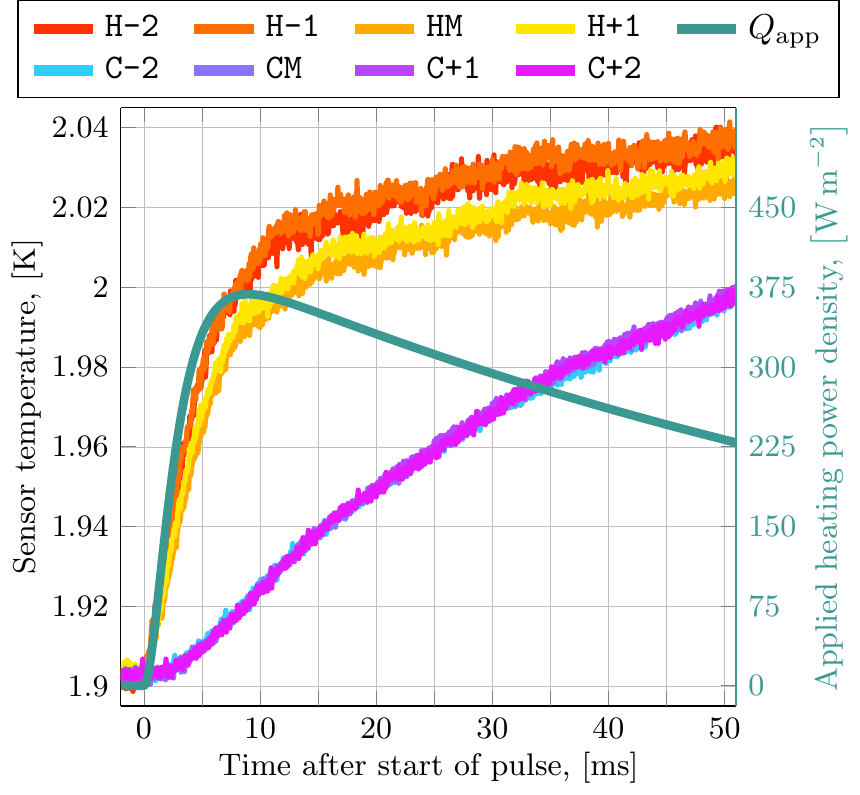}
	\vspace{-15pt}
	\caption{Representative measurement results during the first 60~\ms\ of a slow--pulse test with peak applied heating power density 370~\W. The initial bath temperature was 1.9~\K, and it did not change during the test.}
	\label{fig:slowPulse_firstPlot}
\end{figure}

		Figure \ref{fig:slowPulse_firstPlot} shows a representative measurement of the thermal transient during the first 50~\ms\ of a slow--pulse test in the closed channel. The bath temperature during the test was 1.9~\K. 

During the transient, all sensor temperatures on the heater remain close to each other, reaching about 2~\K\ after 10~\ms, and 2.03~\K\ after 50~\ms. The temperature variation between heater sensors is about $\pm$5~\mK\ after 50~\ms, which is comparable to the estimated measurement uncertainty from Table \ref{tab:uncertainty}. The kink in slope during the temperature rise coincides with the peak in the applied heating power density.

The temperatures in the channel are essentially indistinguishable during the entire test, meaning the helium in the channel heats up uniformly, reaching 1.996~\K\ after 50~\ms. The temperature variation between sensors is about $\pm$2~\mK. Despite the pin--holes at either end of the channel, a thermal gradient along the channel is not expected. As a simple estimate, we consider the cooling power of one pin--hole where the channel end is at 1.996~\K, and the bath end at 1.9~\K. We assume the pin--hole helium behaves according to the Gorter--Mellink heat transfer regime (see Section \ref{sec:heatTransferInHelium} for background and Section \ref{sec:HeliumModel} for discussion of such a modelling approach to our setup). At 1.95~\K, the thermal conductivity function of He~II (in Equation \ref{eq:gorterMellinkHeatFlux_withVariableExponent}) is $f^{-1}$ = \sci{6.25}{14}~\QGM. At 4~mm length, $\Delta T/\Delta x$ = 24~\unit{\kelvin\per\metre}. The heat flux through a pin--hole is then 57~\kW. With a pin--hole cross section of \sci{3.41}{-8}~\mm, the cooling power from two pin holes becomes 3.9~\unit{\milli\watt}. The applied heating power peaks around 167~\unit{\milli\watt}, or nearly two orders of magnitude above the pin--hole cooling power. For all our tests, the pin--hole cooling only matters during the long thermal relaxation time after no more energy is supplied to the heater strip.

\subsection{Simulation Using Measured Helium Temperature as $\mathbf{T_\text{ref}}$} \label{sec:HeRefSim}
\begin{figure}[t!]
		\centering
		\includegraphics{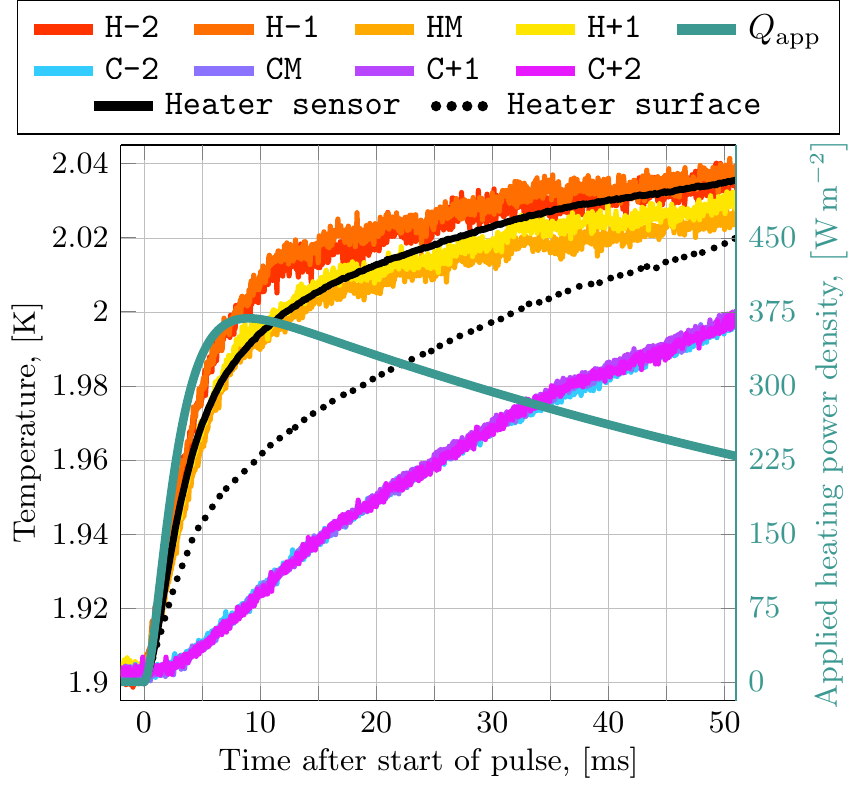}
		\vspace{-15pt}
		\caption{Same slow--pulse test as that shown in Figure \ref{fig:slowPulse_firstPlot}, together with the result of a simulation that models only the material stack. The helium wetted surface of the heater is cooled by the Kapitza expression using the steady state Kapitza parameters found in our Open Bath Paper; \aK\ = 1316.8~\aKUnit\ and \nK\ = 2.528.}
		\label{fig:slowPulse_withHeRefSim}
		\vspace{-10pt}
	\end{figure}

	Figure \ref{fig:slowPulse_withHeRefSim} shows the same test as Figure \ref{fig:slowPulse_firstPlot}, but includes a simulation like the one we used in the Open Bath Paper, namely the time--dependent heat equation with the Kapitza expression (Equation~\eqref{eq:kapitza}) used as a boundary condition at the surface of the heater. 
	
	The solid black curve shows the simulated temperature in the sensor location, and should be compared with the measured heater sensor temperatures. In the simulation shown we have taken the average of the measured helium temperatures, and use this as the reference temperature in the Kapitza expression. This means we do not simulate any helium behaviour.
	
	For the simulation, we have used the steady state Kapitza parameters determined in our Open Bath Paper; \aK\ = 1316.8~\aKUnit, and \nK\ = 2.528. In all plots that show a single test with simulation results, \legendEntry{\texttt{Heater sensor}} refers to the simulated temperature at the location of the sensitive part of the \cx\ sensor within the simulated domain. \legendEntry{\texttt{Heater surface}} refers to the simulated temperature at the Kapitza interface on the heater side.
	
	The simulated sensor temperature is in excellent agreement with measurements. This means, so long as the helium temperature behaviour is known, and the losses are at least as slow as these pulses (reaching peak power after around 9~\ms), the Kapitza heat transfer expression can be expected to capture the thermal behaviour of the system. This result is also a confirmation that even though the Kapitza expression was developed for a situation where the reference temperature was taken far from the heater, it describes the heat transfer also in a shallow channel where the helium is decidedly not far from the heater.
	
	\begin{figure}[t!]
		\centering
		\includegraphics{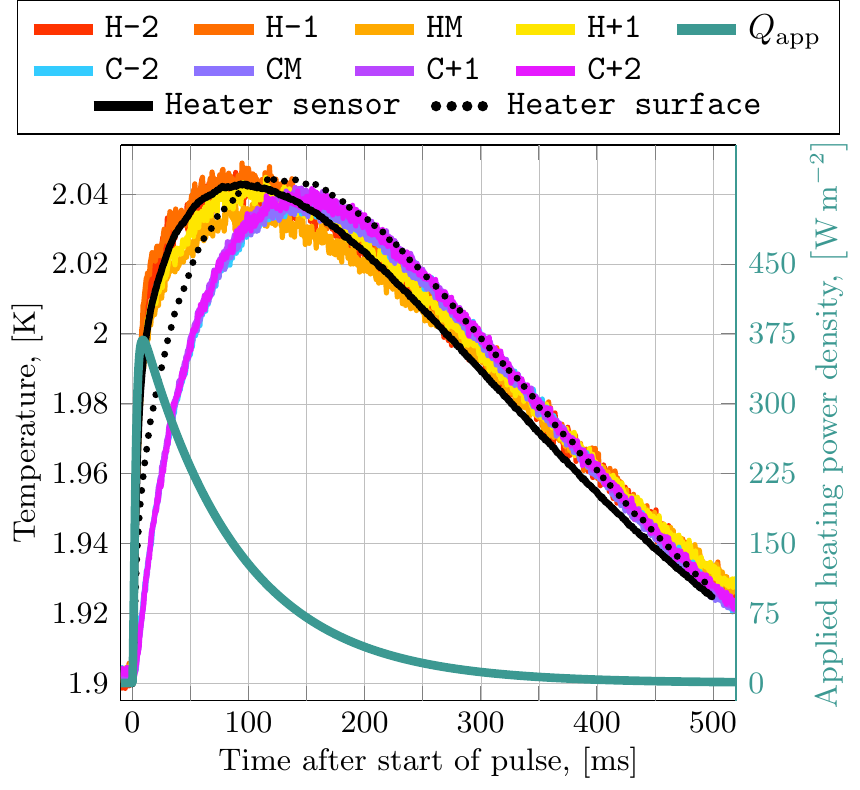}
		\vspace{-15pt}
		\caption{Same simulation result as that in Figure \ref{fig:slowPulse_withHeRefSim}, showing the measurement and simulation result up to 500~\ms.}
		\label{fig:slowPulse_withHeRefSim_long}
		\vspace{-10pt}
	\end{figure}

	\subsubsection{Long Time Scale}	
		In fact, the simulation shown in Figure \ref{fig:slowPulse_withHeRefSim} can be extended to the entire slow pulse, and the result, shown in Figure \ref{fig:slowPulse_withHeRefSim_long}, remains in agreement with the measured heater sensor temperatures all the way up to between 150 and 200~\ms\ after turning on the pulse. The small deviation between measurement and simulation that arises on this very long time scale could stem from the real geometry being three--dimensional, while we simulate it as a simplified one--dimensional geometry, neglecting, for instance, the impact of the PEEK plates themselves.
		
		During the entire test we see the measured temperatures, both on the heater and channel side, remain nearly uniform, with the heater sensor temperatures peaking after about 90~\ms\ at about 2.04~\K. The channel helium temperatures continue rising for another 50~\ms\ after this, as heat keeps transferring from the heater to the helium. The channel temperatures peak at about 2.038~\K. The simulation result, with the heater surface temperature rising above the sensor temperature after about 100~\ms, shows that during the slow relaxation back to the bath temperature, heat transfer along the sensor leads contributes more to cooling than cooling to helium. Since the measured sensor temperatures are above the simulation result at long time scales, it seems our simplified model slightly overestimates the heat flux along the sensor leads. Note that the discrepancy between simulation and measurement is no more than 6~\mK, which is just a little more than the estimated measurement uncertainty.
	
	\subsubsection{Short Time Scale}
\begin{figure}[ht]
			\centering
			\includegraphics{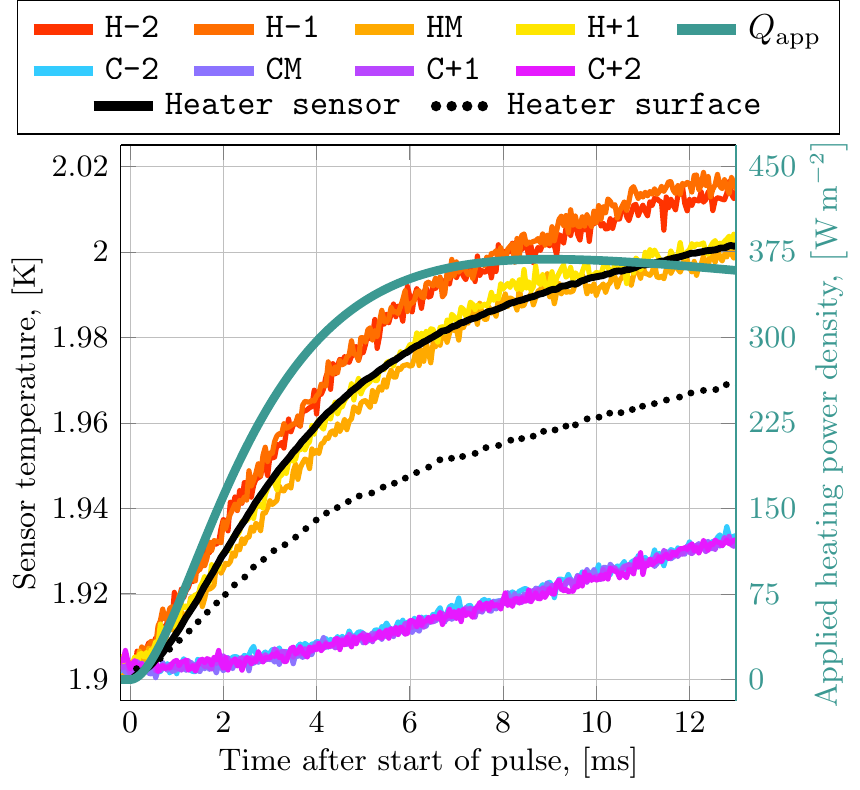}
			\vspace{-15pt}
			\caption{Same simulation result as that in Figure \ref{fig:slowPulse_withHeRefSim}, showing the measurement and simulation result up to 13~\ms.}
			\label{fig:slowPulse_withHeRefSim_short}
		\end{figure}

		Figure \ref{fig:slowPulse_withHeRefSim_short} shows the same simulation result as in figures \ref{fig:slowPulse_withHeRefSim} and \ref{fig:slowPulse_withHeRefSim_long}, highlighting the first 12~\ms.
		
		All temperatures, both simulated and measured, rise smoothly from their initial values. The heater sensors rise roughly linearly between 1 and 4~\ms, corresponding to the region where the applied heating power density rises linearly. After this, the slope tapers off, both due to the pulse nearing its peak, and because the heat capacity of the materials in the stack rises as the temperature grows. The helium channel temperatures start rising very slowly at first, since the heat transfer to the channel depends on the surface temperature of the heater. It then settles into a linear rise starting around 6~\ms. From the simulation we see that the linear region in the measured helium temperatures corresponds to the heater surface temperature having reached a point where the temperature difference between heater and helium is roughly constant, so a nearly steady heat flux crosses the heater--helium interface. This time window also corresponds to the region where the applied heat pulse remains nearly constant around 360~\W\ between about 6 and 13~\ms.
		
		We conclude that the relatively simple one--dimensional material stack model, with the Kapitza expression as the boundary condition representing helium cooling, using the measured channel helium temperature as modelling input, allows for simulation of the transient response to a slow pulse with excellent accuracy all the way from the start of this particular pulse up to around 150~\ms.
	
\subsection{Increasing Peak Power}
\begin{figure}[ht]
		\centering
		\includegraphics{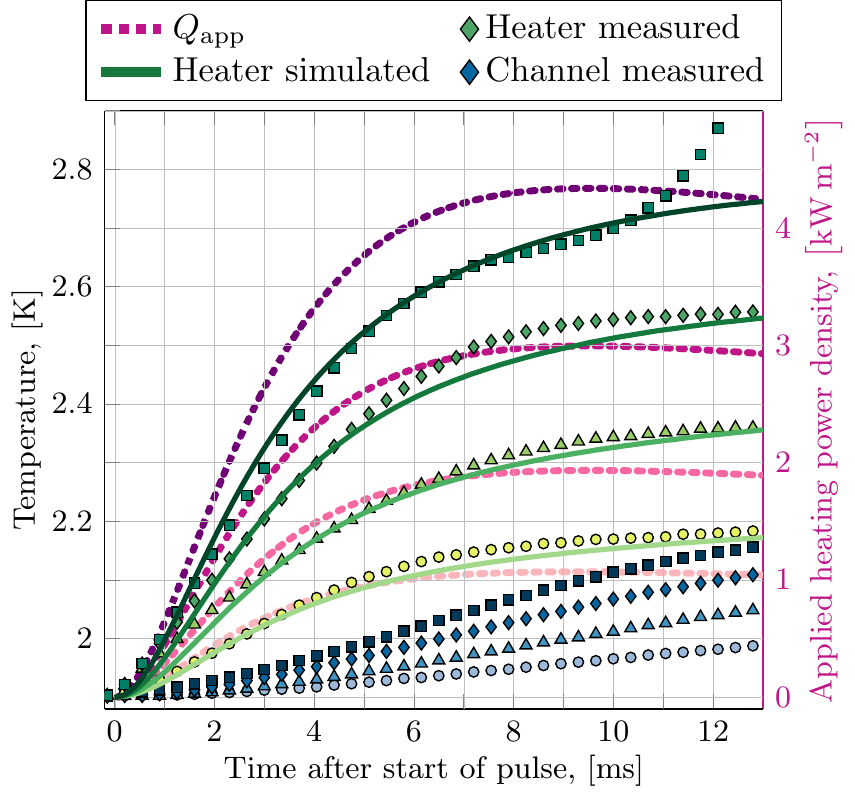}
		\vspace{-15pt}
		\caption{Measurement and simulation results for four successively larger slow pulses with peak applied heating power densities of 1.07, 1.93, 3.00, and 4.34~\kW. The peaks all occur after between 9.2 and 9.5~\ms. Curves of measured temperature data are the average of the four sensors on either the heater or the channel side respectively. Only every seventh data point is plotted. Dashed curves in pink and purple hues are the applied heating power densities: \legendEntry{Qapp}. Symbols in green hues are heater sensor measurement data: \legendEntry{Heater measured}. Symbols in blue hues are channel sensor measurement data: \legendEntry{Channel measured}. Solid curves in green hues are the simulated heater sensor results: \legendEntry{Heater simulated}.}
		\label{fig:severalSlowPulsesTogetherWithHeRefSims}
	\end{figure}

	Figure \ref{fig:severalSlowPulsesTogetherWithHeRefSims} shows measured and simulated temperatures for four successively larger slow pulses. The measured temperatures are represented as the average temperature among the four heater or four channel mounted sensors at each point in time. The average heater sensor temperatures are shown in green hues, while channel temperatures are shown in blues. The dashed curves in purple hues are the applied heating power density for each pulse. The solid curves in greens are the simulated temperatures at the heater sensor location based on the same kind of simulation presented in Section \ref{sec:HeRefSim} using the plotted applied heating power densities.
	
	For the weakest pulse that peaks at 1.07~\kW, the average channel temperature reaches 1.965~\K\ after 10~\ms, with a variation around this average of $\pm$5~\mK. The average heater sensor temperature has risen to 2.170~\K, with a variation of $\pm$20~\mK. For the strongest pulse that peaks at 4.34~\kW, the average channel temperature reaches 2.113~\K, with a variation of $\pm$13~\mK. The average heater sensor temperature reaches 2.700~\K, with a variation of $\pm$65~\mK.
	
	After 10~\ms\ for the largest pulse, the channel temperature approaches \Tl, and rather than flatten out, the slope of the heater sensor temperature increases. The heat transfer characteristics at the helium--heater interface completely change once He~II transitions to He~I. Rather than heat being transmitted by the highly efficient Kapitza conductance into superfluid helium, a natural convection heat transfer regime takes over. The applied heating power still needs to cross the interface, but for the less efficient convection regime, the heater temperature must rise considerably to transfer the same power. It is this regime change that is seen as the rapid temperature rise starting at 10~\ms.
	
	All slow--pulse tests with peak applied heating power densities above 0.68~\kW\ see helium in the channel reach, and eventually go above, \Tl. We never see the channel temperature go above 4.2~\K, however, even for the largest peak applied heating power density (peak at 4.34~\kW). Note that the simulations in Figure \ref{fig:severalSlowPulsesTogetherWithHeRefSims} do not include any attempt at modelling the details of the regime change.
	
	From the excellent agreement between simulation results and measured heater sensor temperatures, it is clear that using the Kapitza heat transfer expression as a boundary condition in transient modelling works well across a wide range of slow--pulse amplitudes, so long as the helium temperature to use as reference in the Kapitza expression is known. We expect the modelling to work also in open bath situations for these kinds of pulses, where the helium temperature does not change during the loss event. 
	
	During the simulations shown in Figure \ref{fig:severalSlowPulsesTogetherWithHeRefSims} we also estimate the fraction of applied heating power that flows backwards to the bath along the sensor leads of the heater sensors rather than forwards across the heater--helium interface (Kapitza interface). At the very start of the pulse, no heat flows across the Kapitza interface, since the heater temperature must first rise appreciably above the helium temperature. So, during the first 20~\us, most of the applied heat flows backwards. After this time, a larger and larger fraction is transferred across the Kapitza interface, until after around 200~\us, the heat flow fractions have stabilised to where only about 8\%\ flows backwards. This fraction slowly decays, remaining above about 6\%\ during the time window shown in the figure. From the steady state measurements in the Open Bath Paper we found the heat leaks backwards represent about 2 to 3\%\ of the input power, so during the transient it is to be expected that somewhat more energy flows into the material stack to heat it up.

	\section{Fast Pulses in Applied Heating Power Density}
\begin{figure}[t!]
	\centering
	\begin{subfigure}[b]{\columnwidth}
		\centering
		\includegraphics{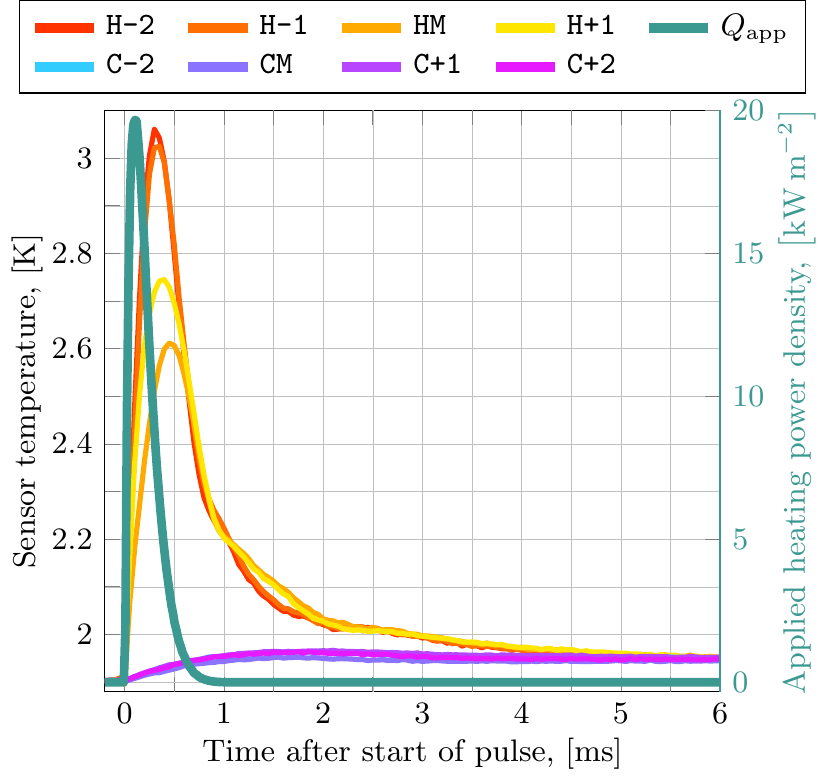}
		\vspace{-5pt}
		\caption{Focus on heater sensors during the first 6~\ms\ of the test.}
		\label{fig:representativeFastPulse_heaterSide}
	\end{subfigure}
	
	\begin{subfigure}[b]{\columnwidth}
		\centering
		\includegraphics{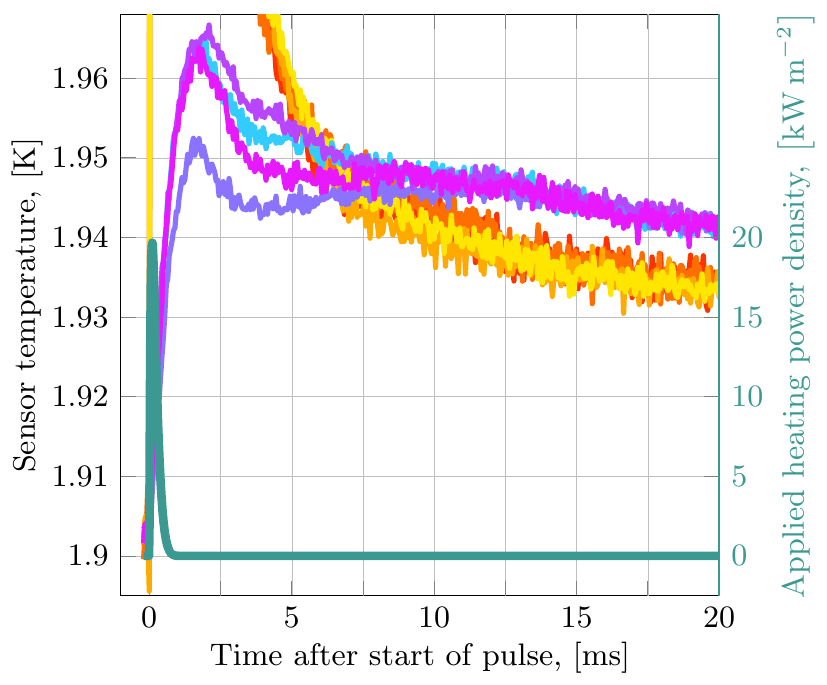}
		\vspace{-5pt}
		\caption{Focus on channel sensors during the first 20~\ms\ of the test.}
		\label{fig:representativeFastPulse_channelSide}
	\end{subfigure}
	\vspace{-15pt}
	\caption{Measurements from a fast--pulse test. Peak applied heating power density is 19.7~\kW, 100~\us\ after the start of the pulse. Initial temperature was 1.9~\K, and no bath temperature change was seen during the test.}
	\label{fig:representativeFastPulse}
\end{figure}

		Figure \ref{fig:representativeFastPulse} shows measured temperatures during a fast--pulse test that reached a peak in applied heating power density of 19.7~\kW\ after 100~\us, with Figure \ref{fig:representativeFastPulse_heaterSide} highlighting the heater sensors, and Figure \ref{fig:representativeFastPulse_channelSide} the channel sensors and the early stages of the thermal relaxation to bath temperature. 

The larger applied heating power density means that although the fraction of heat flow backwards through the material stack is similar to that seen during slow--pulse tests, the absolute heat flow is larger. This in turn helps explain the larger variation between heater sensor peak temperatures during the test; there are variations in the region around the heater sensors (as discussed in the Open Bath Paper), and for a larger heat flow, these geometrical variations cause larger thermal gradients.

The peak heater sensor temperatures occur after between 300 and 450~\us, with \sensor{H-2} and \sensor{H-1} being faster, while \sensor{HM} is slowest. Recall that the data acquisition frequency is 20~\kHz, so each measurement point is 50~\us\ apart. The faster sensors also reach the higher temperatures, consistent with there being slightly more material to heat up between the heater strip and the sensor itself, which reduces the peak temperature during a finite energy pulse such as this. Note that, with a single exception, all sensors have their peak temperatures happen before the applied heating power density has fallen to 15\%\ of its amplitude. The exception is the \sensor{HM} sensor during the strongest fast pulse, which has its peak temperature when the applied heating power density is about 2\%\ of its amplitude.

The channel helium heats up much slower than the heater strip; this is because heat transferred into the helium must first build up in the heater in order to establish the necessary temperature difference across the Kapitza interface before then heating the helium. This, of course, requires time to develop due to the heat capacity of the heater strip steel. Note that helium temperatures start rising immediately, meaning heat transfer across the interface, though it takes time to fully develop, influences the energy distribution among the regions of the system right away. 

Helium temperatures peak roughly at the same time, 1.7~\ms\ after the start of the pulse, and only the middle channel sensor appears to behave differently from the others, showing a peak temperature about 13~\mK\ below the other three channel sensors. After between 5 and 7.5~\ms, the heater sensor temperature falls below the channel temperature. After 10~\ms\ an equilibrium is established, with a quasi--stable temperature difference between the heater and channel sensors. After this time, the two groups of sensors show no significant internal variation, and they slowly relax to the bath temperature over about 150~\ms. The temperature difference between the two groups slowly narrows until all temperatures end up at 1.9~\K. During this slow relaxation the heater surface and the channel helium are in thermal equilibrium, with only a very small heat flow across the interface as heat leaks out of the channel helium.

Note that the complicated state between the peak helium temperature and the point where the smooth thermal relaxation sets in grows longer for stronger pulses. As the peak applied heating power density goes above about 36~\kW\ the state lasts until 20~\ms\ after the start of the pulse. By 54.1~\kW, it lasts till 30~\ms, while for the strongest pulse, at 128.9~\kW, it lasts until 300~\ms\ after the pulse. The time at which the heater sensor temperatures fall enough to reach the channel temperatures is consistently never longer than 10~\ms. 

During the entire fast pulse shown, the total energy developed in the heater strip is about 2.65~\mJ. For a simple estimate, we assume all this energy goes into heating only the stainless steel heater strip. The heater strip volume is about \sci{2.25}{-8}~\mmm, and at 2.7~\K, its heat capacity is 79550~\heatCap. For the given energy input, the heater temperature would rise to 3.4~\K. On the channel side, we have about \sci{5.58}{-8}~\mmm\ of helium, which at 1.95~\K\ has a heat capacity around 660~000~\heatCap. Assuming the energy developed in the heater strip at some point in time is all contained in the helium, its temperature should rise to about 1.972~\K, just 10 to 20~\mK\ above the measured peak helium temperatures. These two simple estimates, that do not consider heat transfer from the heater to the helium, heat leaks to the materials below the heater, or heat flow out of the helium, are clearly in line with the observed values.

\subsection{Large Fast Pulses} \label{sec:largeFastPulses}
\begin{figure}[ht]
		\centering
		\includegraphics{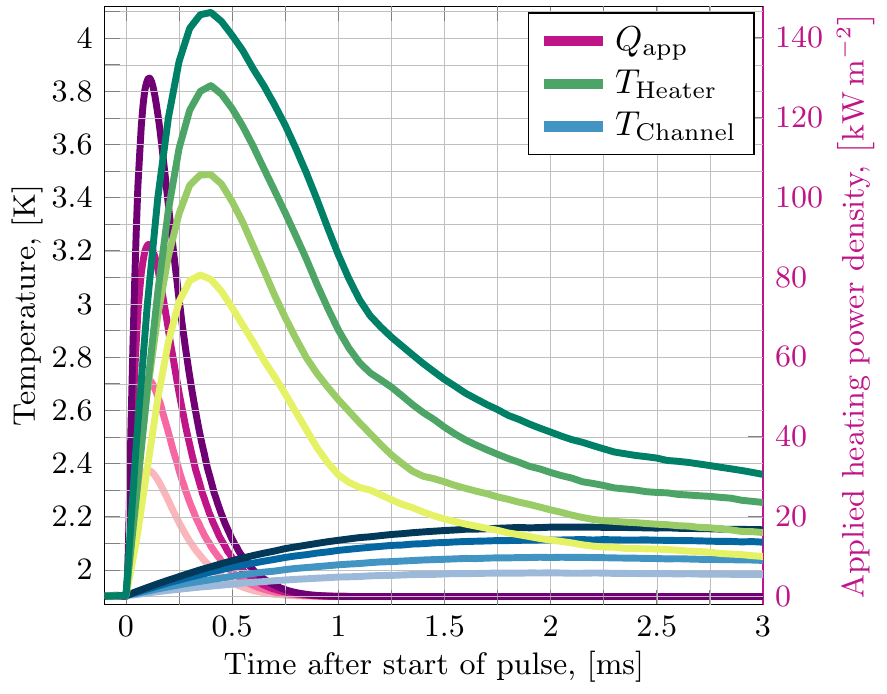}
		\vspace{-15pt}
		\caption{Measurement of four successively larger fast pulses with peak applied heating power densities of 31.5, 54.1, 88.2, and 129.8~\kW. The peaks all occur after between 100 and 110~\us. Curves of measured temperature data are the average of the four sensors on either the heater or the channel side respectively.}
		\label{fig:severalFastPulsesTogether}
	\end{figure}

	Figure \ref{fig:severalFastPulsesTogether} shows four successively larger fast--pulse tests, focusing on the first 3~\ms. The measurement data is represented by the average temperature for each group of sensors; heater (green hues) and channel (blue hues) respectively. Table \ref{tab:fastPulsesSummary} summarises the features of the figure. As from Figure \ref{fig:representativeFastPulse}, during these stronger fast--pulse tests, sensors \sensor{H-2} and \sensor{H-1} give very similar results, with peak temperatures at the upper end of the variation given in Table \ref{tab:fastPulsesSummary}, while \sensor{HM} has both the lowest amplitude and latest time of peak. This remains consistent with the thermal path between heater and sensor being somewhat longer for this sensor than the others. The times of peak heater sensor temperature are consistently around 350~\us, sensor \sensor{HM} being the outlier. 
	
	On the channel side, the peak temperatures always happen after around 2.1~\ms, with no clear tendency for one particular sensor to deviate from the others. Furthermore, the peak helium channel temperatures remain very similar for all pulses. We observe no clear consistency in which sensor measures the lowest peak, though the middle channel sensor gives the lowest temperature in about half the fast--pulse tests. 
	
	\begin{table}[ht]
		\centering
		\caption{Summary of fast--pulse test from Figure \ref{fig:severalFastPulsesTogether}. \Qapp\ is here the peak value during each test. $T_\text{heater}^\ast$ and $T_\text{channel}^\ast$ are the peak values from the figure, while $\Delta T_\text{heater}$ and $\Delta T_\text{channel}$ are the variations around the peak value. $\tau_\text{heater}^\ast$ and $\tau_\text{channel}^\ast$ are the time of peak temperature, while $\Delta\tau_\text{heater}$ and $\Delta\tau_\text{channel}$ are the variations around this time.}
		\begin{tabular}{l | l l l l | l}
			$Q_\text{app}$				& 31.5 				& 54.1 					& 88.2 				& 129.8 				& \kW \\ \hline
			
			$T_\text{heater}^\ast$		& 3.109 			& 3.486					& 3.822				& 4.096 				& \K \\
			$\Delta T_\text{heater}$	& $_{-250}^{+280}$	& $_{-450}^{+370}$		& $_{-610}^{+480}$	& $_{-640}^{+610}$ 		& \mK \\
			$\tau_\text{heater}^\ast$	& 350 				& 400					& 400				& 400 					& \us \\
			$\Delta\tau_\text{heater}$	& $_{-50}^{+100}$	& $_{-50}^{+100}$		& $_{-50}^{+150}$	& $_{-50}^{+300}$ 		& \us \\[8pt]
			
			$T_\text{channel}^\ast$		& 1.989 			& 2.048					& 2.114				& 2.161 				& \K \\
			$\Delta T_\text{channel}$	& $_{-7}^{+7}$		& $_{-8}^{+6}$			& $_{-3}^{+4}$		& $_{-9}^{+10}$ 		& \mK \\
			$\tau_\text{channel}^\ast$	& 2.0 				& 1.95					& 2.15				& 2.15 					& \ms \\
			$\Delta\tau_\text{channel}$	& $_{-0.3}^{+0.4}$	& $_{-0.05}^{+0.15}$	& $_{-0.0}^{+0.3}$	& $_{-0.05}^{+0.15}$ 	& \ms
		\end{tabular}
		\label{tab:fastPulsesSummary}
	\end{table}

	Note that up until the strongest pulses, with peak applied heating power densities above about 100~\kW, the temperature variation within the channel remains around $\pm$7.5~\mK, which is only barely larger than the measurement uncertainty of about $\pm$5~\mK. This means the channel maintains the 120~\um\ depth evenly along its length, otherwise we would see larger variation between the peak temperatures. A significant depth variation would mean the sensors touch effectively different volumes of helium, but these volumes are subject to very similar heat flows from the heater. This conclusion only holds if we can assume that regions of the channel about 30~mm apart (sensor--to--sensor distance) are effectively isolated from each other on the time--scale investigated here. To check this, we assume heat transfer within the channel helium behaves according to the Gorter--Mellink regime. For a worst--case scenario, we take the thermal conductivity function value in Equation \eqref{eq:gorterMellinkHeatFlux_withVariableExponent} at its peak; \sci{6.6}{14}~\QGM. For the typical peak temperature variation of $\Delta T$ = 15~\mK, and a distance $\Delta x$ = 30~mm, we get a heat flux of 18.6~\kW, meaning a total power of 6.92~\mW. Consider a situation where this power flows from a 30~mm long section of the channel at 2.115~\K\ to an adjacent section 15~\mK\ cooler. Taking the heat capacity at 2.1~\K\ as \sci{1.1}{6}~\heatCap, and disregarding the temperature increase in the cooler section, it would take about 27~\ms\ before the surplus energy in the hotter region is transferred to the colder one. This is an order of magnitude longer than the time--scale we are looking at. Considering also that the thermal conductivity function of helium falls quite rapidly once above about 1.95~\K, while the heat capacity grows towards \Tl, it is safe to regard the helium channel as a set of isolated volumes on the time--scale of a few milliseconds.
	
	For the strongest pulse we see the channel temperatures reach very close to \Tl. The calibration uncertainty is large enough that even the highest channel sensor temperature is not conclusively above \Tl. For the slow--pulse tests, we clearly see the heat transfer regime change in the heater sensor data as the channel reaches \Tl, but no such evidence is seen during fast pulses. An important difference is that for fast pulses the peak channel temperature happens well after the applied heating power density has settled back to zero, at which point the only driver for heat flow across the helium--heater interface is the energy stored in the heat capacity of the heater strip (and material stack behind it). We will revisit this question in Section \ref{sec:regimeChangeFastPulses}.
	
\subsection{Simulating Fast Pulses With Measured $\mathbf{T_\text{ref}}$} \label{sec:fast_HeRefSim}
\begin{figure}[t!]
		\centering
		\includegraphics{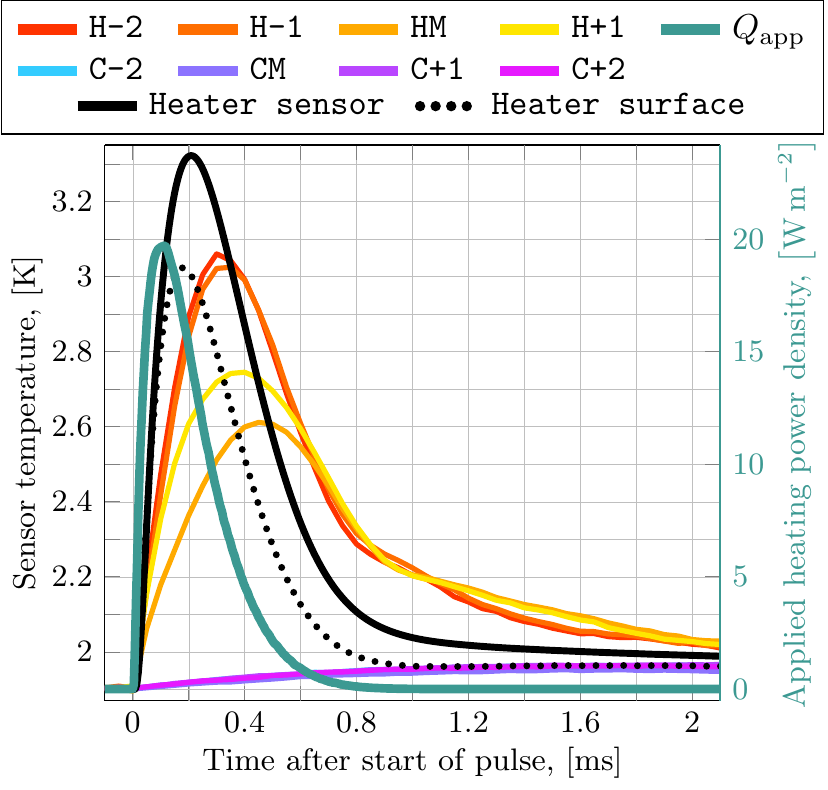}
		\vspace{-5pt}
		\caption{Same fast--pulse test as that shown in Figure \ref{fig:representativeFastPulse}, together with the result of a simulation that models only the material stack (exactly like that for slow pulses in Section \ref{sec:HeRefSim}), which uses the measured helium channel temperature as the reference in the Kapitza cooling boundary condition.}
		\label{fig:fastPulse_withHeRefSim}
	\end{figure}

	Figure \ref{fig:fastPulse_withHeRefSim} shows the same test as that in Figure \ref{fig:representativeFastPulse} together with the result of a simulation like that successfully used on slow pulses in Section \ref{sec:HeRefSim}. While slow pulses could be simulated with excellent accuracy using the measured channel helium temperature, it is clear that fast pulses cannot. Comparing the measured heater sensor temperature curves to the solid black simulated sensor temperature we find that the simulated temperature rises significantly faster than the measured temperatures. This is seen also on the simulated heater surface temperature which represents the temperature of the heater at the Kapitza interface.
	
	The simulated heater surface temperature peaks after 177~\us\ at 3.023~\K\, and the simulated heater sensor temperature peaks after about 210~\us\ at 3.322~\K. The measured heater sensor temperatures peak after around 350~\us, and about 0.5~\K\ lower. For stronger pulses, the peak simulated sensor temperature occurs slightly later; for the strongest fast pulse the peak happens after 240~\us, which is still much sooner than the 400~\us\ seen in measurements. There is also a considerable difference in the peak temperature value in Figure \ref{fig:fastPulse_withHeRefSim}, where the simulation reaches about 0.5~\K\ higher than the average of the measured peak temperatures. At least some of this discrepancy can be explained by the variations in materials surrounding the sensor. We investigate the effect of dimensional and material parameter variation in Section \ref{sec:slowingDownSimulations}. 
	
	The simulated thermal relaxation after the peak is also different from measurements; after around 0.8~\ms\ both simulation and measurement transitions to a slower cooling trajectory, but measurements are still at around 2.3~\K\ and remain higher than the simulated trajectory which has fallen to about 2.1~\K\ before the clear slope change. The simulated transient predicts the heater surface temperature should reach the helium temperature after about 1~\ms. This is very consistent between all tested fast pulses, with simulation of the strongest fast pulse taking till 1.2~\ms. When the heater surface temperature is equal to the helium temperature, the Kapitza expression gives zero heat flux across the interface. However, the channel helium temperature keeps rising until 1.7~\ms\ after the start of the pulse (rising until 2.1~\ms\ for the strongest pulses). This is a clear sign that simply applying the steady state Kapitza boundary condition at the heater surface does not capture the real physics at this time--scale; if there were no heat transfer from the heater strip, the helium in the channel would not keep heating up.

	\section{Model the Channel Helium} \label{sec:HeliumModel}
\begin{figure}[ht]
	\centering
	\includegraphics{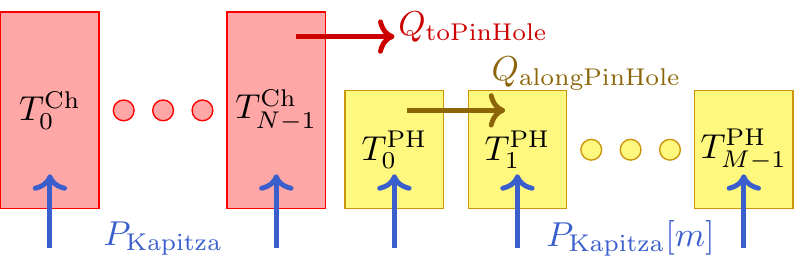}
	\caption{Diagram of channel and pin--hole nodes, and how they interface. There are $N$ channel nodes, denoted by the ``Ch'' superscript, and $M$ pin--hole nodes, denoted by the ``PH'' superscript. From the temperature distribution in the pin--hole we find {\color{DarkGoldenrod4}$Q_\text{alongPinHole}$}, and use this with the cross section of the pin--hole to determine {\color{Red3}$Q_\text{toPinHole}$}. {\color{RoyalBlue3}$P_\text{Kapitza}$} denotes the volumetric heat flow into (or out of) the node from the heater strip. The index {\color{RoyalBlue3}$[m]$} indicates that we use the pin--hole temperature at each node in the Kapitza expression.} 
	\label{fig:channelPinHoleNodes}
\end{figure}

		To develop our modelling of the setup further, we introduce a simple model of the channel helium itself. To account for the pin--holes, we take our simulated domain to be one half of the channel, along its length, connected to one pin--hole at the right end. At the left end of the simulated domain, corresponding to the middle of the channel, we assume zero heat flow. The pin--hole is simulated along its 4~mm length, and we keep the right end, corresponding to the point where it touches the bath, at the bath temperature. Figure~\ref{fig:channelPinHoleNodes} shows a diagram of the boundary and interface nodes of the simulated domain.

The boundary and interface conditions are as follows;
\begin{itemize}
	\itemsep-0.25em 
	\item
	At channel node $T^\text{Ch}_0$ we assume zero heat flux; we take the channel to be symmetrical around the middle.
	\item
	At pin--hole node $T^\text{PH}_{M-1}$ we assume the helium temperature is equal to the bath temperature.
	\item 
	At pin--hole node $T^\text{PH}_{0}$ we assume the helium temperature is equal to the temperature of channel node $T^\text{Ch}_{N-1}$.
	\item 
	At channel node $T^\text{Ch}_{N-1}$ we let flow the heat flux {\color{Red3}$Q_\text{toPinHole}$}. We find this heat flux from the estimated heat flux that flows between the two first pin--hole nodes, {\color{DarkGoldenrod4}$Q_\text{alongPinHole}$}, and scale this by the ratio between the pin--hole cross section and the channel cross section.
	\item 
	At each node, a volumetric heat input/output {\color{RoyalBlue3}$P_\text{Kapitza}$} is found from the Kapitza expression, and either added as an energy source or subtracted as an energy sink.
\end{itemize}

In the simulated domain itself, we assume the Gorter--Mellink regime is always dominant, and discretise Equation~\eqref{eq:transientHeII}, following a similar approach to Fuzier (see Eq. 5.22 in Ref. \cite{fuzierThesis}). We obtain the following numerical scheme;
\begin{equation} \label{eq:1DHelium_discretised}
	\begin{split}
		C_{i} \frac{T_{i}^{n+1} - T_{i}^n}{\Delta t} &= \frac{1}{2\Delta x} \left[ \mathcal{A}_{i} \frac{T_{i+1}^{n+1} - T_{i}^{n+1}}{\Delta x} - \mathcal{B}_{i} \frac{T_{i}^{n+1} - T_{i-1}^{n+1}}{\Delta x} \right]
		\\
		&\:+ \frac{1}{2\Delta x} \left[ \mathcal{A}_{i} \frac{T_{i+1}^{n} - T_{i}^{n}}{\Delta x} - \mathcal{B}_{i} \frac{T_{i}^{n} - T_{i-1}^{n}}{\Delta x} \right] \\
		&\:+ P_i^n
	\end{split}
\end{equation}
where $C_{i}$ is the volumetric heat capacity of helium at node $i$; $T_{i}^n$ = $T(x_i, t_n)$; $\Delta x$ denotes the grid spacing in the $x$--direction; $P_{i}^n$ is a time-- and space--dependent volumetric heat source/sink, and;
\begin{equation} \label{eq:AB_helium}
	\begin{split}
		\mathcal{A}_{i} &= \left( \frac{1}{f\left( \frac{1}{2}\left[T_{i}^n + T_{i+1}^n \right] \right)} \right)^{\textstyle\frac{1}{m}} \left( {\frac{\Delta x}{\abs{T_{i+1}^n - T_{i}^n} + \epsilon}} \right)^{\textstyle\frac{m-1}{m}},\\
		\mathcal{B}_{i} &= \left( \frac{1}{f\left( \frac{1}{2}\left[T_{i}^n + T_{i-1}^n \right] \right)} \right)^{\textstyle\frac{1}{m}} \left( {\frac{\Delta x}{\abs{T_{i}^n - T_{i-1}^n} + \epsilon}} \right)^{\textstyle\frac{m-1}{m}},
	\end{split}
\end{equation}
where $\epsilon$ is a small number, here \sci{1}{-8}, to ensure numerical stability when the thermal gradient is very close to zero, and $m$ = 3.4.

The heater strip runs along the bottom of both the channel and the pin--holes, and to simulate the heating of the channel/pin--hole we use the same material stack simulation as that in sections \ref{sec:HeRefSim} and \ref{sec:fast_HeRefSim}. The key difference is that now, instead of using the measured helium temperature as input to the Kapitza boundary condition on the heater surface of the material stack we use the simulated channel helium temperature. To simplify the simulation load we run only a single material stack simulation, getting only a single heater surface temperature, and use this to find the total volumetric energy flow into/out of each node of the channel/pin--hole helium. This simplification works well because the heat flux out of the pin--hole, cooling the channel, is not large enough to cause any meaningful thermal gradient along the length of the channel.

\begin{figure}[ht]
	\centering
	\includegraphics{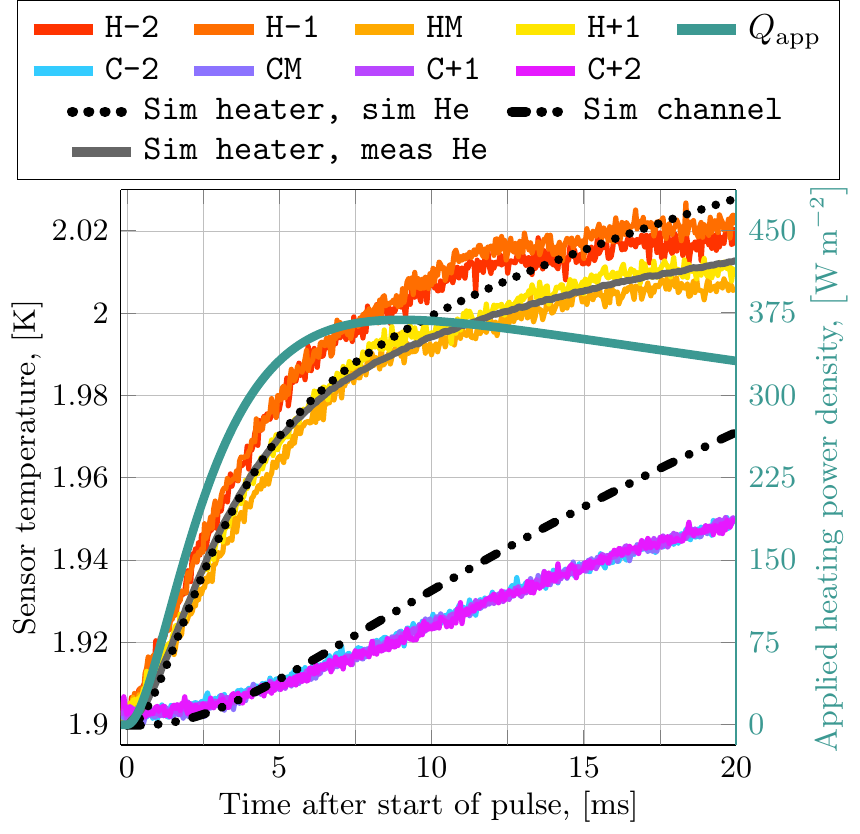}
	\vspace{-15pt}
	\caption{Slow--pulse test with the simulation result also shown in Figure~\ref{fig:slowPulse_withHeRefSim_short}, plotted until 20~\ms, together with the simulation result that includes simulation of channel and pin--hole helium. \legendEntry{\texttt{Sim heater, sim He}} denotes the simulated heater sensor temperature from the new approach, and \legendEntry{\texttt{Sim channel}} denotes the simulated channel temperature. \legendEntry{\texttt{Sim heater, meas He}} is the old simulation result obtained by using the measured helium channel temperature. Note that the initial temperature of the two simulations differ by about 2.5~\mK.}
	\label{fig:compareSlowHeRefSimAndHeSim}
\end{figure}

To assess the validity of this model, we test it on the slow pulse for which we know the simulation works well when using the measured helium temperature. Figure~\ref{fig:compareSlowHeRefSimAndHeSim} shows a comparison between this new simulation approach, that also simulates the helium, and the previous simulation approach where we used the measured channel helium temperature. The solid grey curve is the same as the result shown in Figure~\ref{fig:slowPulse_withHeRefSim_short}, plotted up to 20~\ms. The black dotted curve is the simulated heater sensor temperature from the new approach, while the black dash--dot--dotted curve is the helium temperature from the new approach.

Looking at the simulated heater temperature, we see there is no difference between the two results before around 5~\ms, and only after around 10~\ms\ do the trajectories clearly deviate. The simulated helium temperature, however, appears to follow a different trajectory than the measured channel temperature immediately, though even after 5~\ms\ it is still within the estimated calibration uncertainty of $\pm$5~\mK. That the deviation between the two simulation approaches only becomes clear around 10~\ms\ is expected; the Kapitza expression, with the \nK\ exponent on both the heater and helium temperatures, gives very similar heat flux values for helium temperatures within a few millikelvin of each other. After 5~\ms, the simulated heater surface temperature from both simulation approaches is about 1.943~\K, with the new approach that also simulates the helium being on a slightly steeper trajectory. The measured helium temperature is around 1.9095~\K, while the simulated helium temperature is 1.911~\K. The Kapitza expression, using \aK\ = 1316.8~\aKUnit\ and \nK\ = 2.528, gives 304~\W\ for the lower, measured helium temperature, and 290~\W\ for the higher, simulated helium temperature. This means the Kapitza cooling heat flux is only 5\%\ larger with the measured helium temperature.

The reason the simulated helium temperature deviates from the measured temperature is more involved, but also expected; we make the channel by pressing two PEEK plates together with a total 16 aluminium bolts. While having so many bolts means the the clamping force is distributed along the plates, the joining of separate parts, necessarily, leaves a gap where the two plates meet. We have placed Kapton tape between the two plates, which helps form a better seal nearest the bolts where the clamping force is greatest. There will still remain small gaps from the manufacturing tolerances of parts which will be filled with helium; some of these gaps will be in contact with the channel helium. This additional helium will, of course, serve to increase the effective heat capacity of the channel. In addition to this, we do not account for heat transfer paths going from the channel helium out to the bath through the PEEK that surrounds it. We conclude that neglecting these parasitic cooling effects is valid until 5~\ms.

\subsection{Fast Pulses With Helium Simulation} 
	\begin{figure}[t!]
		\centering
		\begin{subfigure}{\columnwidth}
			\centering
			\includegraphics{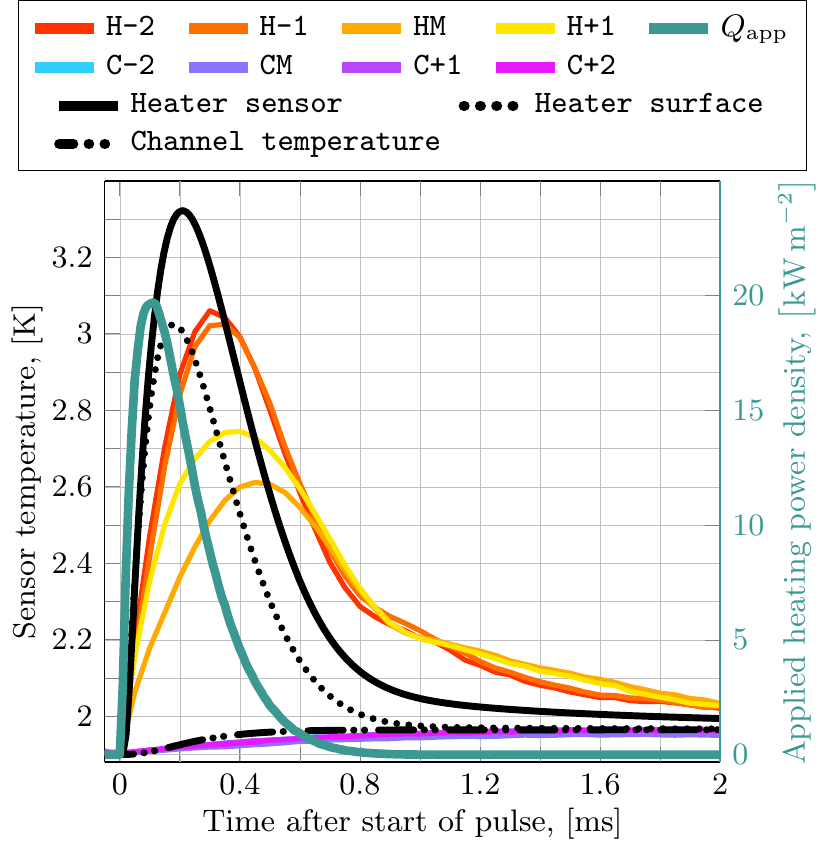}
			\vspace{-5pt}
			\caption{Focus on heater sensors during the first 2~\ms\ of the test.}
			\label{fig:representativeFastPulse_withHeSim_heaterSide}
		\end{subfigure}
		
		\begin{subfigure}{\columnwidth}
			\centering
			\includegraphics{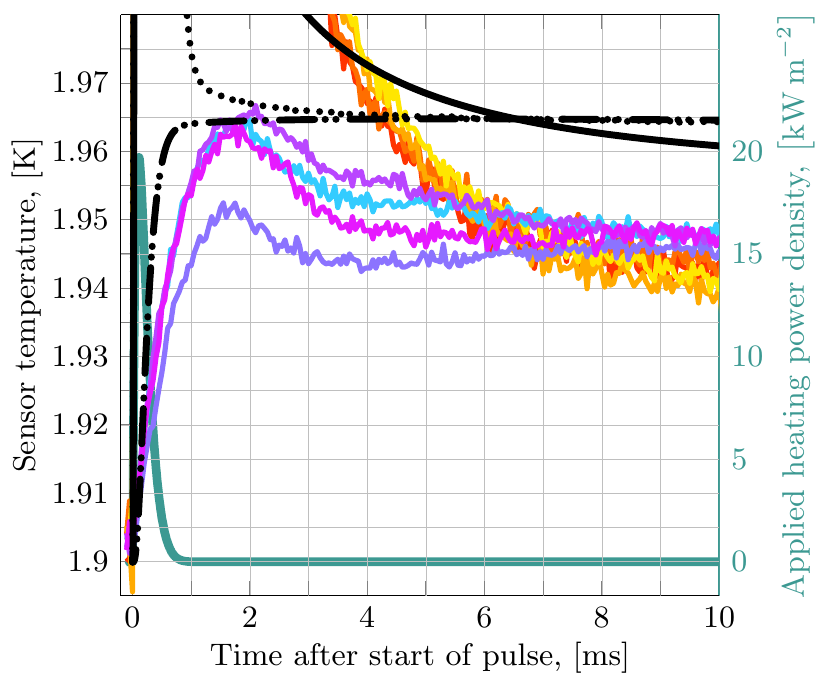}
			\vspace{-5pt}
			\caption{Focus on channel sensors during the first 10~\ms\ of the test.}
			\label{fig:representativeFastPulse_withHeSim_channelSide}
		\end{subfigure}
		\vspace{-5pt}
		\caption{Measurement from the same fast--pulse test shown in Figure~\ref{fig:representativeFastPulse}, together with simulation results that include the channel helium connected to a pin--hole.}
		\label{fig:representativeFastPulse_withHeSim}
	\end{figure}

	Figure~\ref{fig:representativeFastPulse_withHeSim} shows the result of the new simulation approach compared with the measurements from the same fas--pulse test shown back in Figure~\ref{fig:representativeFastPulse}. The solid black curve is the simulated heater sensor temperature (which should be compared with the measured temperatures). The dotted black curve is the simulated heater surface temperature. The dash--dot--dotted black curve is the simulated helium temperature.

	From Subfigure~\ref{fig:representativeFastPulse_withHeSim_heaterSide}, the simulated heater sensor temperature is not substantially different from the result obtained by using the measured helium temperature (seen in Figure~\ref{fig:fastPulse_withHeRefSim}). The peak temperature is 3.322~\K\ after 209~\us\ with both simulation approaches. The simulated heater surface temperatures, however, differ a little; this new simulation gives the peak simulated heater surface temperature after 179~\us\ at 3.025~\K, which is 2~\us\ after and 2~\mK\ above the simulation that used the measured helium temperature for the Kapitza boundary condition. While not a meaningful difference in terms of our measurement uncertainties, this result is consistent with the different heat flux at the Kapitza interface when looking at the simulated helium temperature in Subfigure~\ref{fig:representativeFastPulse_withHeSim_channelSide}. It rises faster than the measured temperature, meaning that the Kapitza heat flux will be slightly lower during this simulation, which in turn means slightly less effective cooling, leading to the higher peak temperature at a later time.

	In Subfigure~\ref{fig:representativeFastPulse_withHeSim_channelSide} we see the simulated helium temperature stabilise at 1.964~\K\ after about 1~\ms. The measured helium temperatures have a peak around 1.961~\K\ (from the average of the four temperatures), but it takes till 1.7~\ms\ before it happens. So, on the one hand, we can accurately simulate the peak helium temperature value, confirming that the channel helium is isolated from the parasitic cooling effects that eventually causes the slow--pulse simulation deviation around 5~\ms\ in Figure~\ref{fig:compareSlowHeRefSimAndHeSim}. On the other hand, the rate of heat transfer to the helium is clearly not captured correctly by applying the steady state Kapitza expression, even if it gives the correct total energy transfer.
	
	Since we do not consider the parasitic cooling effects, and since the pin--holes provide so little cooling power relative to the volume of the channel, the simulated helium temperature only very slowly decays compared with the measured values. Also, the simulated helium temperature keeps rising until around 6.5~\ms\ where the simulated heater surface temperature has cooled down to that of the helium, at which point no significant heat transfer across the Kapitza interface takes place. This crossing point happens at the same time for both simulation and measurement, and we see the match all the way up to peak applied heating power densities around 100~\kW. For larger pulses than this, the simulated helium temperature reaches \Tl, and we look more into this in Section~\ref{sec:regimeChangeFastPulses}.	

\subsection{Slowing Down the Simulated Transient} \label{sec:slowingDownSimulations}
	We now have a model of our system that can accurately simulate the temperature development for a slow pulse up to 5~\ms. Furthermore, in situations where the helium temperature is known, the heater/material stack part of our modelling setup is essentially accurate for the entire duration of a slow pulse. That same model, with or without simulation of the helium, does not work satisfactorily for fast pulses. This corresponds qualitatively to the observations of UFOs in the LHC; the magnet needs more energy to reach the temperature at which it quenches as compared with a model of the magnet that works well for steady state losses. So, at the millisecond time--scale, processes not yet accounted for in the simple model lead to improved cooling performance that prevents magnets from quenching, and we can clearly see this effect in the slower--than--expected heater and helium temperature rises.
	
	Can we explain the measured results by simply changing parameters within the model as we have implemented it? The following discussion compares the fast--pulse test in Figure \ref{fig:representativeFastPulse_withHeSim} with simulations using various parameters. The results are general to all fast pulses regarding the key metrics of time and amplitude of peak.
	
	\subsubsection{Parameter Space From Open Bath Tests}
		In our Open Bath Paper we found a parameter space comprised of the steady state Kapitza parameter variation and uncertainty in dimensions of the various material layers in the material stack used for transient simulations. Within this space, we could successfully simulate the temperature development of all sensors from about 1~\ms\ after a step in applied heating power density. Between the step and the first millisecond, we saw that the measured temperature rise is slower than simulations. This same parameter space cannot explain the closed channel behaviour. By using \aK\ = 1213.6~\aKUnit, \nK\ = 2.86, 41~\um\ of varnish between the heater strip and the \cx\ sensor, 10~\um\ of EPO--TEK for the sensor lead attachments, and 18~mm of copper leads, which represents the lowest temperature curve within the parameter space, we get a simulated sensor temperature that has a peak temperature value of 2.92~\K\ after 210~\us\ for the same fast--pulse test as that shown in Figure \ref{fig:representativeFastPulse_withHeSim}. This temperature amplitude is about 120~\mK\ below the two sensors \sensor{H-2} and \sensor{H-1}. The average sensor temperature measurement for this test has a peak temperature of 2.845~\K\ after 350~\us. So, while the simulated temperature amplitude is better with the parameter changes, they still lead to a peak around 150~\us\ sooner than measurements.
	
		So, within the estimated geometrical parameter space, we can get the right temperature amplitude for sensors \sensor{H-2} and \sensor{H-1}, but not the other two heater sensors. But the time of the simulated peak is always at least 100~\us\ too early.
	
	\subsubsection{Variation in Material Parameters}
		We do not get significant improvement even after allowing material parameter variation of $\pm$20\%\ on both thermal conductivity and heat capacity of the materials in the stack. $\pm$20\%\ represents a pessimistic assumption of the measurement uncertainty of material data found in literature (see Appendix A in the Open Bath Paper). Higher heat capacity, of course, slow down the simulated temperature response. Lowering the thermal conductivity of the varnish between the sensor and heater lowers the temperature amplitude, and increasing the steel thermal conductivity also lowers it. A simulation with 1213.6~\aKUnit\ and \nK\ = 2.86, 41~\um\ of varnish, 10~\um\ of EPO--TEK, and 18~mm of copper leads, where we also increase the thermal conductivity of steel by 20\%, lower that of varnish by 20\%\ and increase all heat capacities of the materials in the stack by 20\%\ gives a peak simulated sensor temperature of 2.700~\K, which is 45~\mK\ below sensor \sensor{H+1}, but the peak is only slowed till 215~\us, much too fast.
	
		Furthermore, none of these changes do anything to change the simulated helium temperature rise significantly. With a higher heat capacity of steel, it takes slightly longer before the Kapitza heat transfer builds up, but even with a 20\%\ increase, the time at which the helium reaches the stable temperature seen in Subfigure \ref{fig:representativeFastPulse_withHeSim_channelSide} remains about 1~\ms.
	
	\subsubsection{Thermal Boundary Resistance}
		In our model of the material stack, we have not so far considered the possible effect of thermal contact resistance between the solids. In particular, a thermal boundary resistance between the heater strip and the varnish layer, or between the varnish layer and the \cx\ sensor would help slow down the temperature rise and the amplitude of the simulated temperature response of the sensors. To assess the effect of such a thermal resistance we include a layer of material between the heater strip and the varnish that has an effective thermal conductivity of just 1\%\ that of the varnish itself. This leads to a simulated sensor temperature peak of just 2.474~\K, which is below even that of sensor \sensor{HM}, but it still happens after only 240~\us, much too soon compared with the measured transient. And, again, this change to the model does not alter the response of the helium.
	
	\subsubsection{Thermal Gradient Across the Channel}
		Since the pin--holes have such a small cooling effect on the channel, a reasonable change to the helium modelling is to instead simulate a channel domain upwards, across the channel, to see if the slow helium sensor response could be due to the time it takes heat to propagate from the helium adjacent to the heater to the helium near the sensor 120~\um\ away. For this kind of simulation we disregard the pin--holes completely, and use the heat flux from the Kapitza expression as a Neumann boundary condition at the heater--adjacent node of the channel helium. At the sensor--adjacent node we assume zero heat flux. With these boundary conditions we implement Equation \eqref{eq:1DHelium_discretised} in the same way we do when simulating helium along the channel length.
		
		We find only a negligible thermal gradient across the channel depth; on the order of 5~\unit{\micro\kelvin} between the heater--adjacent and sensor--adjacent helium. This is essentially independent of which helium node we use as the reference temperature for the Kapitza expression. So, with the Gorter--Mellink heat transfer regime we cannot explain the slow helium temperature rise. Note that if the Gorter--Mellink regime is not dominant, it would be because it has not yet had time to develop fully, meaning the laminar heat transfer regime still matters, which transfers heat even better than the Gorter--Mellink regime, leading to even smaller thermal gradients across the channel.
		
	\subsubsection{Inter--Plate Helium}
		A final avenue we have looked at is to try and include the inter--plate helium that is trapped between the two PEEK plates after bolting shut and sealing with \ecco. This enters the realm of speculation, because we do not have reliable estimates for how wide this inter--plate gap is. We have flatness measurements of the PEEK plates when they are not under tension from the bolts which indicates there is a maximum gap--length of 50~\um. After applying the considerable clamping force of the 16 aluminium bolts the resulting gap is certainly not that large. However, if we assume the inter--plate gap, filled with helium, is uniformly 10~\um\ along the entire length of the channel, and that it runs uninterrupted all the way out to the bolt--holes about 6.5~mm from the edge of the channel, we can define a similar decomposed domain for the inter--plate helium as we do for the pin--holes. This does not yield a significantly slower simulated helium temperature response. However, it does result in a weak tendency of the helium temperature to fall after reaching the same peak temperature as that found for a channel alone. This is because now we allow for some heat to flow out of the channel region, along this extra helium volume. 
		
		Extending the speculative added helium domain even further, to include the helium trapped between the aluminium bolts and the PEEK wall of the hole, we can obtain a simulated helium temperature decays more similarly to the measured values, since there is now effectively even more helium into which heat from the channel can be siphoned off. It does not significantly slow down the initial temperature rise however.
		
		Insofar as an inter--plate gap filled with helium exists, it is unlikely that it can be approximated as a continuous channel of even depth along the entire mating region of the two PEEK plates. Furthermore, that the helium around the aluminium bolts should be in direct contact with this hypothetical inter--plate helium is effectively precluded by the fact that where the aluminium bolts clamp down on the two plates, there is also the largest clamping force, meaning the volumes are completely sealed off.
		
%

\subsection{Regime Change During Tests} \label{sec:regimeChangeFastPulses}
	For the strongest applied heating power densities, across all three time--dependent profiles, we see the helium temperature approach and, for some tests, go above \Tl. In order to expand the simulation framework to also include these high--power tests, we must implement some form of regime change once \Tl\ is reached. 
	
	In the helium domain, when a node goes above \Tl\ we change the effective thermal conductivity to that of stationary He~I. This change is done in the thermal conductivity coefficients $\mathcal{A}$ and $\mathcal{B}$ in Equation~\eqref{eq:AB_helium}. 
	
	For the heater--helium interface heat transfer, which is no longer governed by the Kapitza expression from Equation~\eqref{eq:kapitza} when the helium temperature has gone above \Tl, we use a natural convection regime;
	\begin{equation} \label{eq:naturalConvection}
		Q_\text{interface} = Q_\text{NatConv} = a_\text{NatConv} (\Ts - \Tr),
	\end{equation}
	where $a_\text{NatConv}$ is the natural convection surface heat transfer coefficient.
	
	During the relevant time--windows of the high--power tests we see no sign of nucleate boiling or the onset of film boiling, which would be accompanied with chaotic channel temperature measurements as the helium is perturbed by bubbles arising at the heater surface.
	
	Our main interest is not the natural convection heat transfer regime itself, so we only need an order--of--magnitude value for this coefficient. Open bath experiments by Dorey~\cite{Dorey_natConv}, and Mori and Ogata~\cite{MoriOgata_natConv}, indicate it should be in the range 200 to 5000~\unit{\watt\per\square\metre\kelvin}. We use $a_\text{NatConv}$ = 1000~\unit{\watt\per\square\metre\kelvin}. 
	
	\subsubsection{Strongest Fast Pulse}
\begin{figure}[t!]
			\centering
			\begin{subfigure}{\columnwidth}
				\centering
				\includegraphics{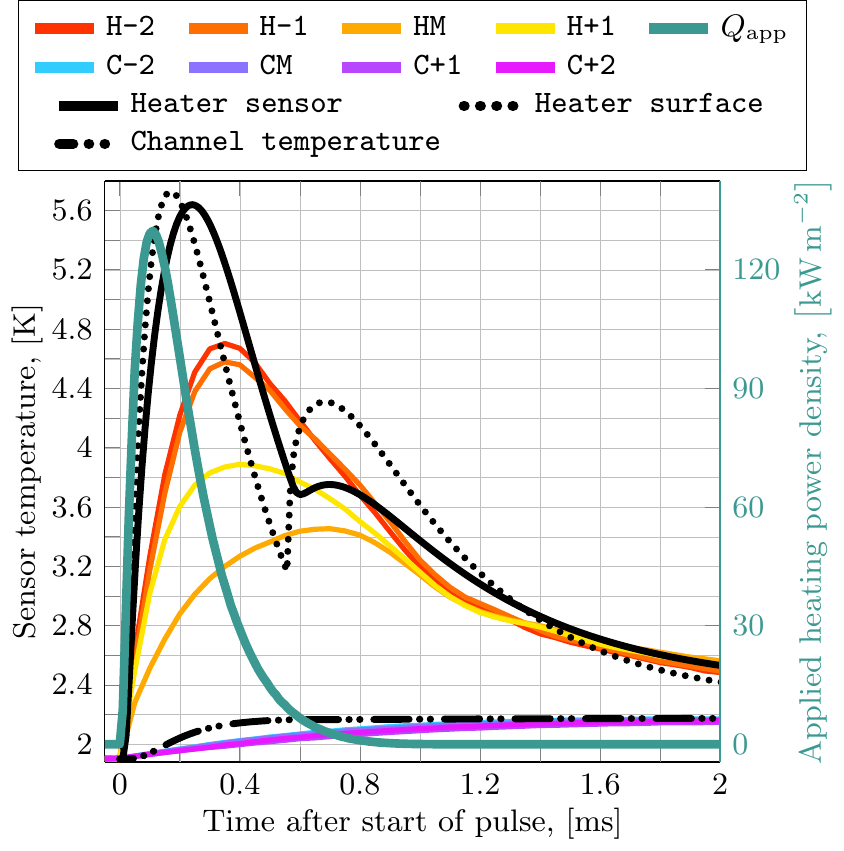}
				\vspace{-5pt}
				\caption{Focus on heater sensors during the first 2~\ms\ of the test.}
				\label{fig:strongestFastPulse_withHeSim_heaterSide}
			\end{subfigure}
			
			\begin{subfigure}{\columnwidth}
				\includegraphics{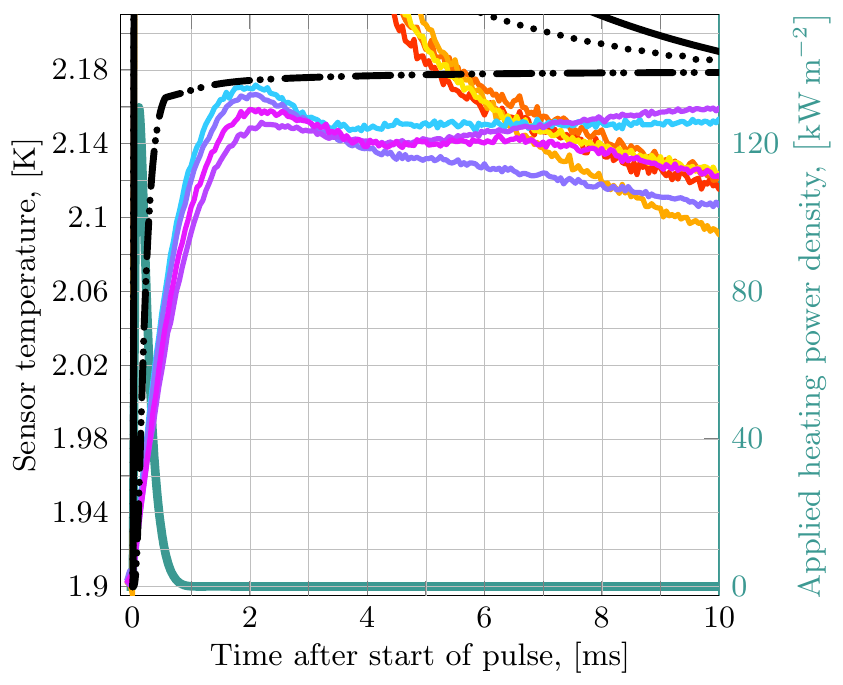}
				\vspace{-5pt}
				\caption{Focus on channel sensors during the first 10~\ms\ of the test.}
				\label{fig:strongestFastPulse_withHeSim_channelSide}
			\end{subfigure}
			\vspace{-5pt}
			\caption{Measurement from the strongest fast--pulse test shown in Figure~\ref{fig:severalFastPulsesTogether}, together with simulation results that include the channel helium connected to a pin--hole.}
			\label{fig:strongestFastPulse_withHeSim}
		\end{figure}

		The strongest fast pulses we have tested are, from our simulations of the channel helium, predicted to see the phase transition to He~I in the channel, with an associated change in heat transfer regime at the heater surface. Since such a regime change is a feature of many models applied to LHC magnets, we look into the strongest fast--pulse test from our measurement campaign. Figure \ref{fig:strongestFastPulse_withHeSim} shows this test together with a simulation where the helium is considered along the length of channel, connected to a pin--hole. For helium above \Tl\ we assume the thermal conductivity of stationary He~I; the channel is horizontal, so we do not expect any lateral convection currents to carry heat.
		
		After about 0.56~\ms, the simulated channel temperature reaches \Tl, seen as a kink in the dash--dot--dotted curve in Subfigure \ref{fig:strongestFastPulse_withHeSim_channelSide}. This is the time we change from Kapitza interface heat transfer to natural convection (Equation~\eqref{eq:naturalConvection}), and where we take the helium thermal conductivity to be that of He~I. This heat transfer regime moves much less heat that Kapitza for the same temperature difference between the heater and helium, and therefore, in Subfigure \ref{fig:strongestFastPulse_withHeSim_heaterSide}, the heater surface temperature shoots up. Immediately before the regime change, the heat flux across the interface, from the Kapitza expression, is 15.2~\kW. After the regime change, the heat flux drops to 1.0~\kW. Energy is still supplied to the heater strip from the external source, contributing to the temperature increase. However, the largest contribution comes from redistribution of the energy already in the heater strip and material stack; during the time between the peak simulated heater temperature and the transition out of the Kapitza regime there is a large thermal gradient across the 50~\um\ thickness of the heater strip, with the bottom of the heater (nearest the sensor) being the hottest point in the simulated domain. This thermal gradient exists in the heater to push the Kapitza heat flux at the interface. When this heat flux suddenly drops to nearly zero, the energy stored in the heater must redistribute itself. This redistribution is seen as the brief period around 0.7~\ms\ where the simulated sensor temperature remains nearly steady at 3.75~\K.
		
		As mentioned in Section~\ref{sec:largeFastPulses}, the measured channel temperatures, for sensors \sensor{C-2} and \sensor{CM} briefly peak above \Tl, though not enough to be beyond the estimated uncertainty range. Looking at the very clear simulated heater sensor feature associated with a heat transfer regime change it appears no such transition actually took place during the test, or, if it did, the Kapitza regime recovered fast enough that we could not observe the excursion.
		
		Note that the apparent agreement between simulation and measurement after around 2~\ms\ in Subfigure~\ref{fig:strongestFastPulse_withHeSim_heaterSide} depends on the choice of $a_\text{NatConv}$ in Equation~\eqref{eq:naturalConvection}, and using 1000~\unit{\watt\per\square\metre\kelvin} does not lead to such agreement for the three weaker fast pulses where the simulations also predict helium temperatures above \Tl. This is, of course, because the natural convection heat transfer coefficient depends on helium temperature, heater temperature, whether the helium flow near the heater is turbulent or laminar, and the confinement of the channel must also play a role, as it changes how the helium flows near the heater as compared with the open bath case for which data is available.

	\section{Comparing Tests}
		
\subsection{Open Bath and Closed Channel Steps}
\begin{figure}[ht]
		\centering
		\includegraphics{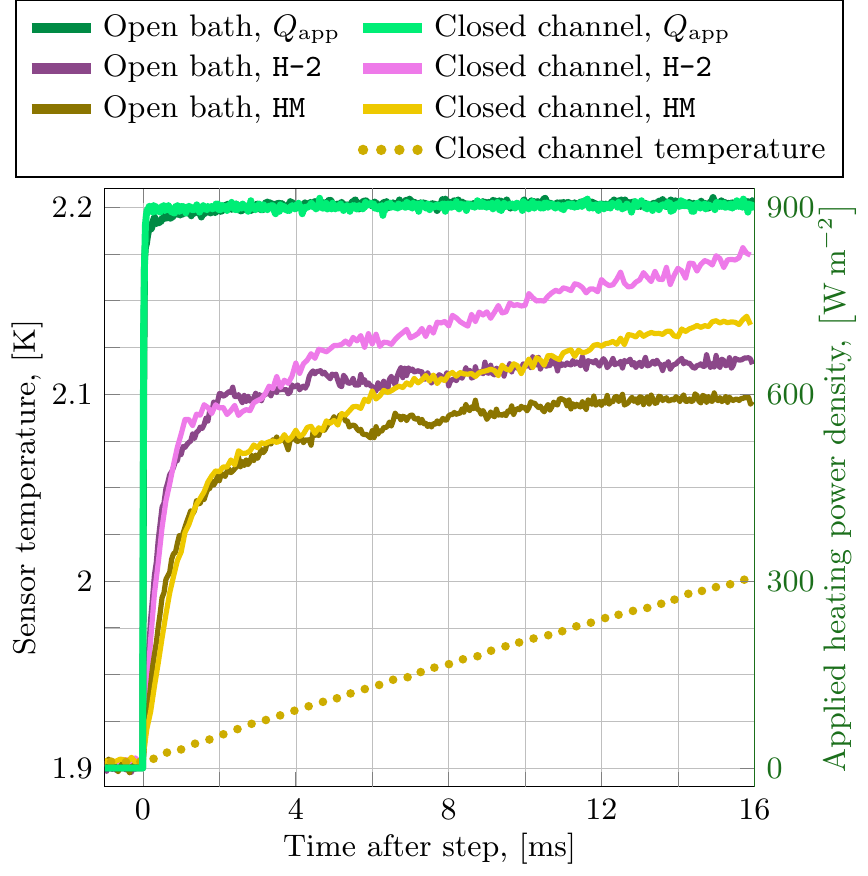}
		\vspace{-15pt}
		\caption{Comparison of step tests that reach around 900~\W\ done in the open bath and closed channel configurations.}
		\label{fig:comparison-Step_43-and-Step_183}
	\end{figure}

	Figure~\ref{fig:comparison-Step_43-and-Step_183} shows two tests; \legendEntry{Open bath} curves refer to a test of a step to 905~\W\ applied to the heater strip cooled by an open bath (test performed as part of the measurement campaign for our Open Bath Paper), while \legendEntry{Closed channel} curves refer to a test of a step to 903~\W\ applied to the heater strip cooled by the closed channel helium. The sensors whose temperatures are plotted are physically the same sensors (called \sensor{D-2} and \sensor{DM} in the Open Bath Paper). The channel temperature is represented by the average temperature of the four channel sensors.
	
	During the first 4~\ms\ there is no distinction between the open bath and closed channel results. That there would be some time window during which no difference is seen makes intuitive sense; heat transferred into the helium from the surface of the heater strip would not immediately know about the helium being confined. There must be some form of time delay, such as a thermal diffusion time, associated with heat transfer within helium, and before the propagating temperature reaches the roof of the channel, the Kapitza interface sees the same as if exposed to an open bath. The question is how long does it take for this information to influence the system. The measurements clearly say around 4~\ms\ from the heater sensor temperatures. The channel temperature measurement, on the other hand, starts rising almost immediately after the step, which is in line with our simulation result that shows no significant thermal gradient across the depth of the channel. Of course, we found some small thermal delay stemming from the material stack that separates the heater surface from the heater sensors, but even within the large parameter space we have tested, this cannot account for more than a few tens of microseconds at most, and certainly not several milliseconds.
	
	The closed channel temperature rise follows a roughly linear trajectory starting at $t=0$, growing by about 6~\unit{\milli\kelvin\per\milli\second}. The temperature growth rate of the closed channel heater sensors after their departure from the open bath results is about 5~\unit{\milli\kelvin\per\milli\second}. That they are not identical is no surprise, since the Kapitza heat transfer expression depends non--linearly on the helium temperature. It is, however, not clear why the closed channel heater temperatures do not appear to grow at all according to the helium temperature during the first 4~\ms. By that time, the channel helium temperature has reached 1.93~\K. Again, 4~\ms\ is much longer than the effective thermal diffusion time through the material stack between the heater surface and the \cx\ sensors.
	
	A similar comparison between identical steps to 546~\W\ show that the open bath heater sensors have an initial rise time around half a millisecond faster than the closed channel equivalent. This is still significantly slower than relevant simulations. The time when the closed channel temperatures start deviating from the open bath ones is still around 4~\ms, however.
	
	This all points towards there being some property of the Kapitza heat transfer, apparent only at this millisecond time--scale, that does not depend intimately on the helium temperature near the interface. Another possibility is that for simulation of heat transfer at this time--scale, the two--fluid nature of He~II cannot be simplified to the Gorter--Mellink heat transfer relation. From our simulations of heat transfer where we use the measured helium temperature as the reference for Kapitza cooling (Section~\ref{sec:fast_HeRefSim}), which isolates the Kapitza interface by using known helium behaviour, we still did not find the right heater behaviour. So not having the right heat transfer relations within the helium is not the only issue. Finally, our model of the region surrounding the heater sensors is simplified to only one dimension, which leaves the small possibility that the real three--dimensional geometry would need to be considered for sufficient accuracy. The main material we neglect when simplifying like this is PEEK, which is essentially a thermal insulator, especially when we consider only the first few milliseconds.
	
\subsection{Tests With Similar Energy Deposition}
\begin{figure}[t!]
		\centering
		\includegraphics{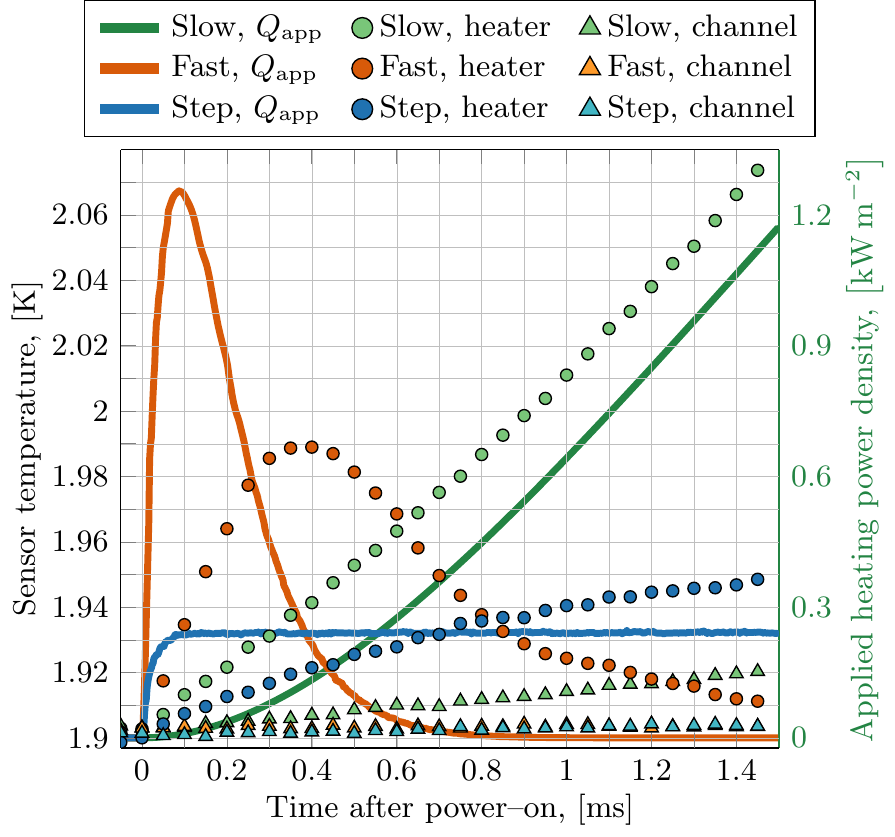}
		\vspace{-15pt}
		\caption{Comparison of three transient tests that have deposited roughly the same amount of energy in the heater strip after 1~\ms. Each temperature curve is the average temperature of the four sensors that comprise the group (heater or channel).}
		\label{fig:compareTestsWithSimilarEnergy}
	\end{figure}

	Figure~\ref{fig:compareTestsWithSimilarEnergy} shows the first 1.5~\ms\ of three different applied heating power density profiles that, after 1~\ms\ have delivered roughly the same amount of energy into the heater strip. The slow pulse has delivered 0.113~\unit{\milli\joule}, the fast pulse 0.153~\unit{\milli\joule}, and the step 0.111~\unit{\milli\joule}. 
	
	The most important observation is that the helium temperature rise during the first millisecond of the slow--pulse test has clearly risen above the initial temperature by about 9~\mK, which is almost twice the estimated calibration uncertainty. The fast--pulse and step tests, on the other hand, after having deposited at least that same amount of energy in the heater strip, see the helium temperature rise by no more than 2~\mK\ which is less than half the calibration uncertainty.
	
	This means that the time--structure of the energy deposition is of fundamental importance to the way energy is distributed between the two main parts of the system; heater and helium. Recall, for the slow pulse shown in the figure, we can very accurately simulate the entire time window shown, including the simulation of the channel helium. That same simulation approach predicts that the channel helium should rise about 3~\mK\ over the first 1~\ms. This is quite similar to the measured rise. However, the simulated heater sensor temperature is faster than the measured temperature, overestimating the Kapitza heat transfer during the first 0.5~\ms. For the fast pulse, simulations give a very different helium behaviour than that seen; the simulated helium temperature rise, like seen before, reaches a flat top value after 1~\ms\ that is 4~\mK\ above the initial temperature, while the measured helium temperature only grows by 1 to 2~\mK\ within this time. Note that the simulation gives a helium temperature rise of 3.5~\mK\ after just 0.5~\ms. The fast--pulse energy deposition is larger than that of the other two tests in the comparison, so the temperature after 1~\ms\ should have been higher than just 1 to 2~\mK\ if all the energy entered the helium. The measured helium temperature rises slowly to a peak about 2 to 3~\mK\ above initial temperature after 8~\ms.
	
	So, for energy depositions that happen on the sub--millisecond time--scale, like fast pulses, an effect we have not accounted for in our modelling, that does not appear to be relevant at longer time--scales, leads to a channel helium temperature rise that is much slower than anticipated. The helium heats up from the heater strip, so the effect must also impact the heat transfer from heater to helium. 
	
	The strongest step performed reached \Qapp\ = 1.02~\kW\ after about 100~\us. It delivered 0.154~\unit{\milli\joule} of energy during the first 330~\us\ of the test. This is the same amount of energy as that delivered during the entire fast--pulse test shown in Figure~\ref{fig:compareTestsWithSimilarEnergy}. During these 330~\us, the heater sensor temperatures are indistinguishable between the two tests. Furthermore, the channel temperature rise during the first 330~\us\ of the step test is about 4~\mK, while the simulated helium temperature rise is about 3~\mK. So, it is only for the fast--pulse tests that it appears the energy transfer into helium is wrong when using the steady state Kapitza expression.
	
	The models used to predict quench levels of LHC magnets incorporate the same physics we have implemented here, except for the internal heat transfer within helium, which is neglected for quench level estimates. In light of the underestimated LHC magnet quench levels during UFO events, we confirm that both heater and helium temperatures remain lower than expected during our controlled experiments. Since all attempts at variations of parameters in our model fail to explain the slower temperature response, on both the heater and helium side, it appears the use of the steady state Kapitza heat transfer expression could be invalid at the UFO time--scale. 
							
	\section{Conclusion}
%
%

To investigate superfluid--helium cooling of a heater subject to time--dependent heating on the millisecond time--scale, as relevant to UFO events in the LHC, we built an experimental setup consisting of a heater and a confined volume of He~II cooling the heater from one side. The setup was previously validated in open bath experiments, where we found steady state Kapitza parameters for the stainless steel heater strip we use. Our setup measures the temperature of the heater strip and the channel helium.

We develop a heat transfer model based on the time--dependent heat equation together with the steady state Kapitza heat transfer expression as a cooling boundary condition at the heater--helium interface. We validate the model against measurements where the heater strip is subjected to a slow pulse in heating power that peaks after 9~\ms\ and lasts for a total of 400 to 500~\ms. When using the measured helium channel temperature as the reference temperature in the Kapitza heat transfer expression, we can accurately simulate the first 150~\ms\ of such a pulse.

We expand this model to include the helium volume by considering a one--dimensional helium domain along the heater connected to a pin--hole at one end. We assume heat transfer within helium always adheres to the fully turbulent Gorter--Mellink regime. With this extension of the model, we can accurately simulate the first 5~\ms\ of slow pulses in heating power. The deviations after this are tied to parasitic cooling effects in the helium stemming from the setup being built as two parts pressed together to form the helium channel.

For fast, UFO--like pulses, which deliver peak power after 100~\us\ and last for less than 1~\ms\ in total, we consistently find that both the heater and the channel helium temperatures rise more slowly than what is predicted by the heat transfer model we validated for slow pulses. The slower temperature rise cannot be explained by material parameter variations, changes in the dimensions of material layers in the model, addition of a thermal boundary resistance between the heater and the sensor, a thermal gradient in the helium across the depth of the channel, or the effect of adding a hypothetical volume of helium between the two PEEK mounting plates.

Comparing slow--pulse, fast--pulse, and step tests that deliver the same amount of energy to the heater strip after 1~\ms, we find that for the UFO--like fast--pulse energy deposition, the expected helium temperature rise, as predicted by our modelling, is higher and happens sooner than what we see in the measurements. The simulated helium temperature rise from the other two energy deposition profiles is in line with measurement. There must, therefore, be a missing, but important effect, not accounted for in our model.

Since fast--pulse simulations using the measured helium temperature in the Kapitza cooling expression do not give heater temperatures that agree with measurements, the missing effect is not isolated to the helium alone. As only the Kapitza interface influence both the heater and helium domains simultaneously, this points towards the steady state Kapitza heat transfer expression not being valid at the UFO event time--scale.
		
	\appendix
		\renewcommand{\thesection}{\Alph{section}}
		\numberwithin{equation}{section}
		\numberwithin{figure}{section}
		\renewcommand\thefigure{\thesection.\arabic{figure}} 
		
	\printbibliography
	
\end{document}